\documentstyle[epsfig]{mn}
\begin{document}
\input BoxedEPS.tex
\newcommand{\aip}{{\small ${\cal AIPS}$}}
\newcommand{\gtsim}{\mbox{{\raisebox{-0.4ex}{$\stackrel{>}{{\scriptstyle\sim}}
$}}}}
\newcommand{\ltsim}{\mbox{{\raisebox{-0.4ex}{$\stackrel{<}{{\scriptstyle\sim}}
$}}}}
\newcommand{\s}{$\stackrel{\rm s}{.}$}
\newcommand{\h}{$^{\rm h}$}
\newcommand{\m}{$^{\rm m}$}
\newcommand{\pp}{$\stackrel{\prime\prime}{.}$}
\newcommand{\de}{$^{\circ}$}
\newcommand{\p}{$^{\prime}$}
\newcommand{\arc}{$^{\prime\prime}$}
\newcommand{\marc}{^{\prime\prime}}
\newcommand{\rs}{{\em $r_s$}}
\newcommand{\DPM}{{\em DPM}}
\newcommand{\alf}{{\displaystyle\biggl({\nu_{\rm h} \over \nu_{\rm l}}\biggr)^{\alpha}} }

\newcommand{\figstart}[1]
    { \begin{figure}[htb]
      \begin{picture}(0,#1) }
\newcommand{\figend}[4]
    { \end{picture}
      \special{#1}
      \caption[#2]{#3}
      \label{#4}
      \end{figure} }
\newcommand{\fig}[5]
    { \figstart{#1}
      \figend{#2}{#3}{#4}{#5} }
\newcommand{\bHS}{\beta_{\mbox{\scriptsize HS}}}
\newcommand{\bBF}{\beta_{\mbox{\scriptsize BF}}}
\newcommand{\nT}{\nu_{\mbox{\scriptsize T}}}
\newcommand{\et}{E_{\mbox{\scriptsize T}}}
\newcommand{\nTn}{\nu_{\mbox{\scriptsize Tn}}}
\newcommand{\nTf}{\nu_{\mbox{\scriptsize Tf}}}
\newcommand{\tn}{\tau_{x\mbox{\scriptsize n}}}
\newcommand{\tf}{\tau_{x\mbox{\scriptsize f}}}
\newcommand{\xn}{x_{\mbox{\scriptsize n}}}
\newcommand{\xf}{x_{\mbox{\scriptsize f}}}
\newcommand{\yn}{y_{\mbox{\scriptsize n}}}
\newcommand{\yf}{y_{\mbox{\scriptsize f}}}
\newcommand{\lln}{l_{\mbox{\scriptsize n}}}
\newcommand{\llf}{l_{\mbox{\scriptsize f}}}
\newcommand{\Dn}{f(\Delta_{\mbox{\scriptsize n}})}
\newcommand{\Df}{f(\Delta_{\mbox{\scriptsize f}})}
\newcommand{\B}{\mbox{$B$}}
\newcommand{\Bo}{\mbox{$B$}_{0}}

\SetEPSFDirectory{/scratch/sbgs/figures/hst/}
\SetRokickiEPSFSpecial
\HideDisplacementBoxes

\title[Photometric redshifts in the Hubble Deep Fields]{Photometric redshifts in the Hubble Deep Fields:
evolution of extinction and the star-formation rate.}
\author[Rowan-Robinson M.]{Michael Rowan-Robinson\\
Astrophysics Group, Blackett Laboratory, Imperial College of Science Technology and Medicine,
Prince Consort Road, London SW7 2BW}
\maketitle
\begin{abstract}
Photometric redshifts are studied with a template approach using data from HDF-N and -S .  The problem 
of aliasing in photometric redshift estimates is investigated in some detail and found
not to be a significant problem if at least four photometric bands are available.  The performance
of the approach presented here appears to exceed that of others in the literature.  The rms accuracy
of the photometric redshifts is 9.6 $\%$ in (1+z), with a 1.5 $\%$ chance of a significant alias when 
four or more photometric bands are used.

With reasonable restrictions, it is possible to determine the dust extinction as well as the 
photometric redshift, provided five or more photometric bands are available.  
An important result is that evolution of  $<A_V(z)>$ with redshift is seen, with higher values than
locally at z = 0.5-1.5, and lower values at z $>$ 2.  This is consistent with current models for
the star formation history of the universe.  

Deconvolving the uv-to-ir seds into an old star and young star component allows determination
of $M_*$ and $\dot{M}_*$ for each galaxy, as well as $z_{phot}$ and $A_V$, provided that
infrared photometric bands are available.  The expected trend of b = $M_*/\dot{M}_* t_0$
increasing to the past is seen.  However there is a great deal of scatter in the relation
between b and sed type, showing that the recent star-formation history is not very well
correlated with the long-term history of a galaxy.

The 2800 $\AA$ luminosity function and star-formation rate are calculated for a large sample 
of HDF-N (2490) and HDF-S (28719) galaxies, using photometric redshifts, for the redshift range 0.2-5.  
The star-formation rates agree reasonably well with those from a variety of other uv, $H_{\alpha}$, 
infrared and submillimetre estimates, and with star-formation histories used to model optical, infrared
and submillimetre source-counts.

\end{abstract}


\section{Introduction}

The Hubble Deep Field provided an important breakthrough in our understanding of the evolution 
of galaxies, permitting for the first time an analysis of the star-formation history of the
universe out to redshift 4 ( Madau et al 1996).  Madau et al concluded that the star-formation rate 
was quite strongly peaked at redshift 1-2, and declined sharply at higher redshifts (see also Madau 
et al 1998).  This decline at high redshift was at least partly due to the neglect of extinction by 
dust, which several subsequent studies 
have shown to be an important effect (Rowan-Robinson et al 1997, Meurer et al 1997, 1999, Pettini et al 
1998, Cram 1998, Steidel et al 1999, Adelberger and Steidel 2000, Hopkins et al 2001, Sullivan et al 2001).  
The correction for dust extinction remains highly uncertain, even for nearby galaxies, and few studies of
the star formation history take any account of the expected evolution of the dust opacity in galaxies with
epoch.   Typical estimates of dust correction factors in the ultraviolet are 2-7.  
A new estimate of the average dust extinction in local galaxies has been made by Rowan-Robinson (2003) using 
ultraviolet, blue and far infrared data.  When the contribution of the heavily extinguished starburst component
is subtracted from the far infrared fluxes, the mean value of $A_V$ in the interstellar mediums of 
nearby (V $<$ 5000 $km/s$) galaxies is found to be 0.31, corresponding to $A_{2800}$ = 0.62 for Milky Way dust.  

In this paper I use large samples of galaxies in the Hubble Deep Fields (HDF), North and South, to 
study the evolution of the ultraviolet radiation-density, the dust opacity and hence the star formation history,
using photometric redshift methods.  I first study the issue of aliasing for photometric redshifts using fixed 
template photometric redshift estimates in HDF-N and HDF-S, comparing the results with those of other
photometric redshift codes.  I then investigate the effect
of introducing extinction, characterized by $A_V$.  Finally I also characterize the
evolution of galaxy seds with redshift, using a new approach which separates the effects
of young stars ($< 10^9$ yrs old) and older stars.  As might be expected the introduction
of additional free parameters leads to increased aliasing, but also gives new insight 
into the evolution of dust extinction with redshift, and into the interpretation of the
Hubble sequence.

The resulting photometric redshift catalogues are used to determine the uv luminosity function
at a range of redshifts from 0.2 -5, the evolution of dust extinction and star-formation rate with redshift, 
and to estimate the stellar mass, $M_*$, and star formation rate, $\dot{M_*}$, in each galaxy.

A Hubble constant of 100 $km/s/Mpc$ is used throughout.

\section{Photometric redshifts in the HDF}
The publication of the HST data on the HDF was an enormous stimulus to work on photometric redshift
derivation.  Lanzetta et al (1996), Mobasher et al (1996), and Gwyn and Hartwick (1996) used
photometric redshift methods to assess the redshift distribution and properties of HDF galaxies.  
Madau et al (1996) used Lyman drop-out galaxies in the Hubble Deep Field to 
derive estimates for the star formation rate at z = 2 - 4.5.
This analysis appeared to demonstrate that the star formation drops off steeply at z $>$ 2.  
Connolly et al (1997) used photometric redshifts for the brighter HDF galaxies
to derive the star formation rate at z = 0.5 - 2.  Their method involved the
use of J, H and K data in addition to the HST U,B,V,I data to determine the 
redshifts.  Although this certainly improves the reliability of the
photometric redshifts, it does restrict the analysis to galaxies with J $<$ 23.5, ie
a small fraction of the galaxies ($< 10 \%$) detected in the HDF by HST.  
Fernando-Soto et al (1999) derived photometric redshifts
for a larger sample of galaxies detected in 7 bands (UBVIJHK) and were able
to test their predictions against the spectroscopic redshift catalogue of Cohen et al (2000).
Pascarelle et al (1998) used the Lanzetta et al (1996) photometric redshifts to study the
ultraviolet luminosity-density as a function of redshift.  Lanzetta et al (2002) use their
photometric redshifts to study the evolution of the surface brightness distribution function
in galaxies.  Arnouts et al (1999) have used photometric redshifts in HDF-N
to study the redshift evolution of clustering and Teplitz et al (2001) have carried out a similar
study with a much large sample of galaxies in a 0.5 square degree field centred on HDF-S.  
Steidel et al (1999) have reanalyzed the HDF
Lyman drop-out galaxies in HDF-N, to compare the results with their own Lyman drop-out
surveys.  Thompson et al (2001) used photometric redshifts to study the star-formation history
using NICMOS data in HDF-N.  Chen et al (2003) used photometric redshifts with the Las Campanas 
H-band survey to 
study the evolution of the rest-frame B-band luminosity function.  A comparison of different 
photometric redshift methods was made by Hogg et al (1998).  

All these studies used a set of fixed templates to determine photometric redshifts.
Mobasher and Mazzei (1998) and Le Borgne and Rocca-Volmerange (2002) have used realistic, evolving 
galaxy seds to determine photometric redshifts.

Bolzonella et al (2000) have used simulations to study the effects of different choices 
of spectral energy distribution (sed) and photometric bands on the accuracy of photometric 
redshifts, and have included the effect of extinction as a free parameter.

\begin{figure}
\epsfig{file=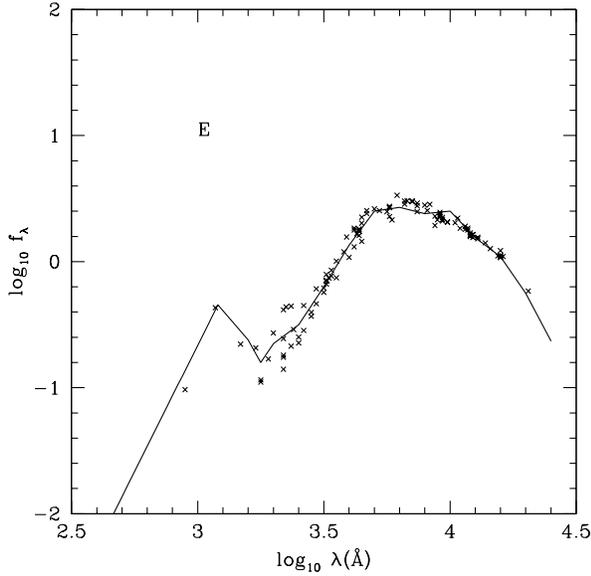,angle=0,width=8cm}
\caption{Assumed sed for elliptical galaxies (solid curve), with photometric data plotted for galaxies with 
spectroscopic redshifts (crosses).}
\end{figure}

\begin{figure}
\epsfig{file=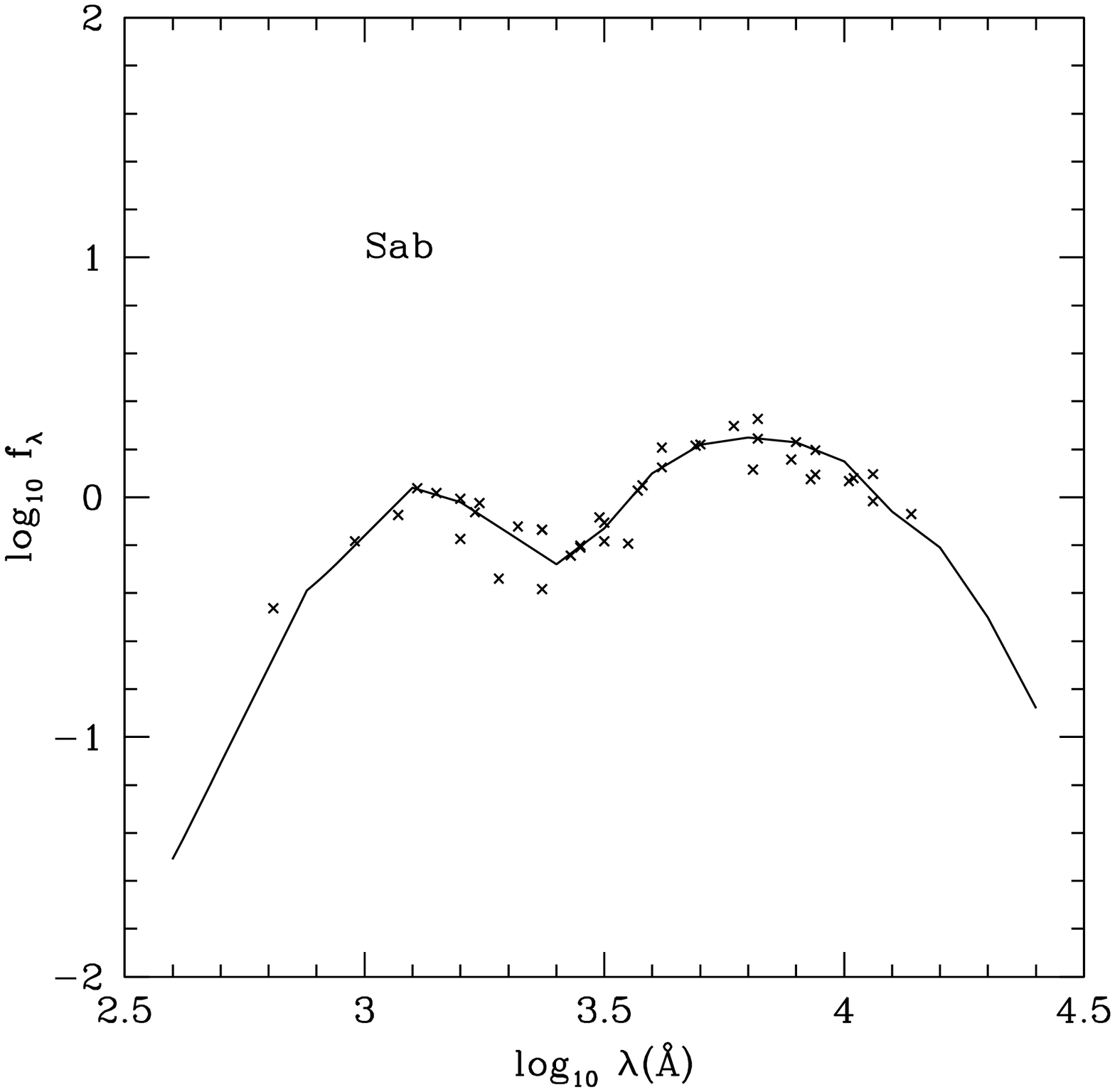,angle=0,width=8cm}
\caption{Assumed sed for Sab galaxies, with data for galaxies with spectroscopic redshifts.}
\end{figure}

\begin{figure}
\epsfig{file=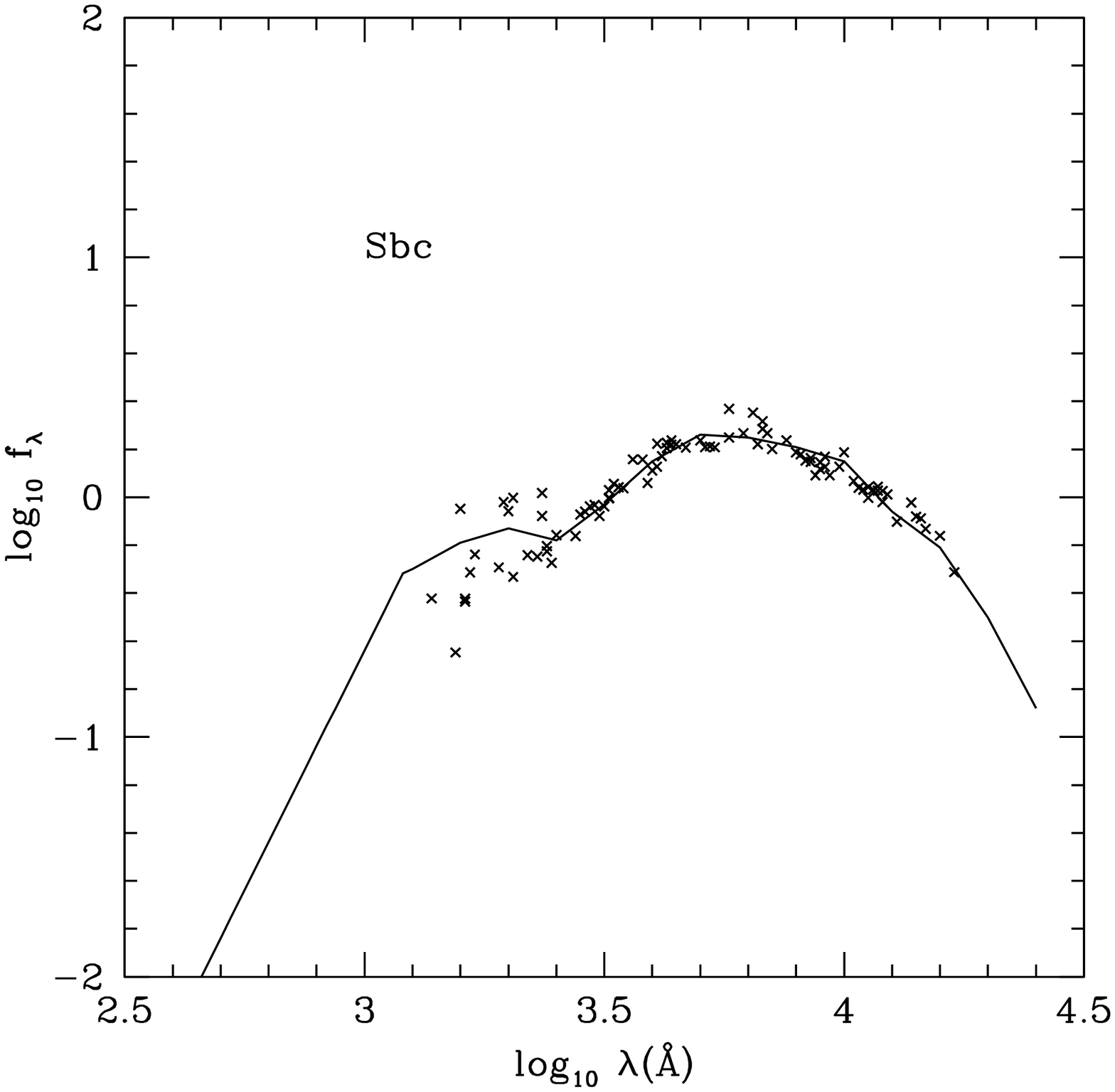,angle=0,width=8cm}
\caption{Assumed sed for Sbc galaxies, with data for galaxies with spectroscopic redshifts.}
\end{figure}

\begin{figure}
\epsfig{file=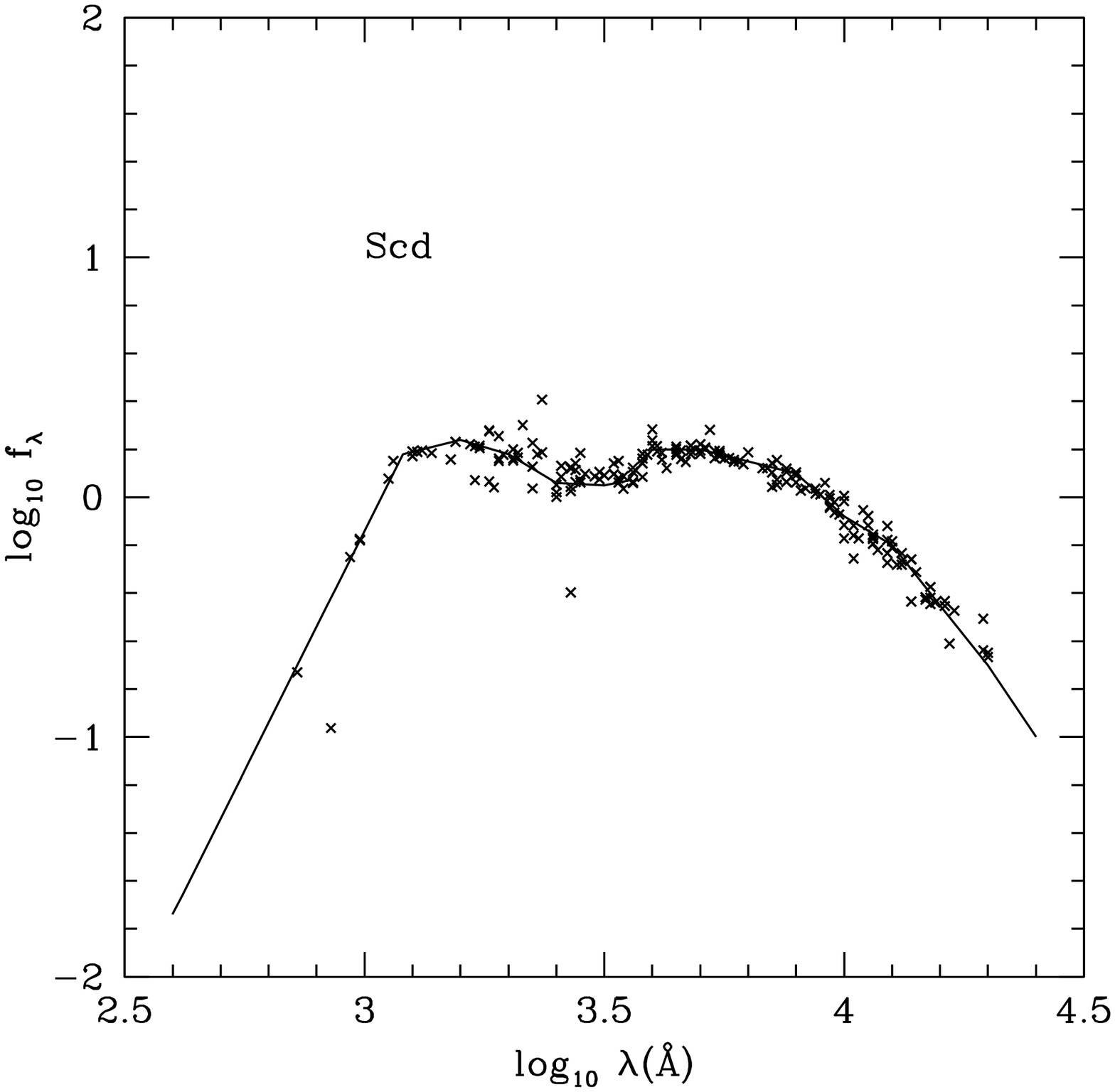,angle=0,width=8cm}
\caption{Assumed sed for Scd galaxies, with data for galaxies with spectroscopic redshifts.}
\end{figure}

\begin{figure}
\epsfig{file=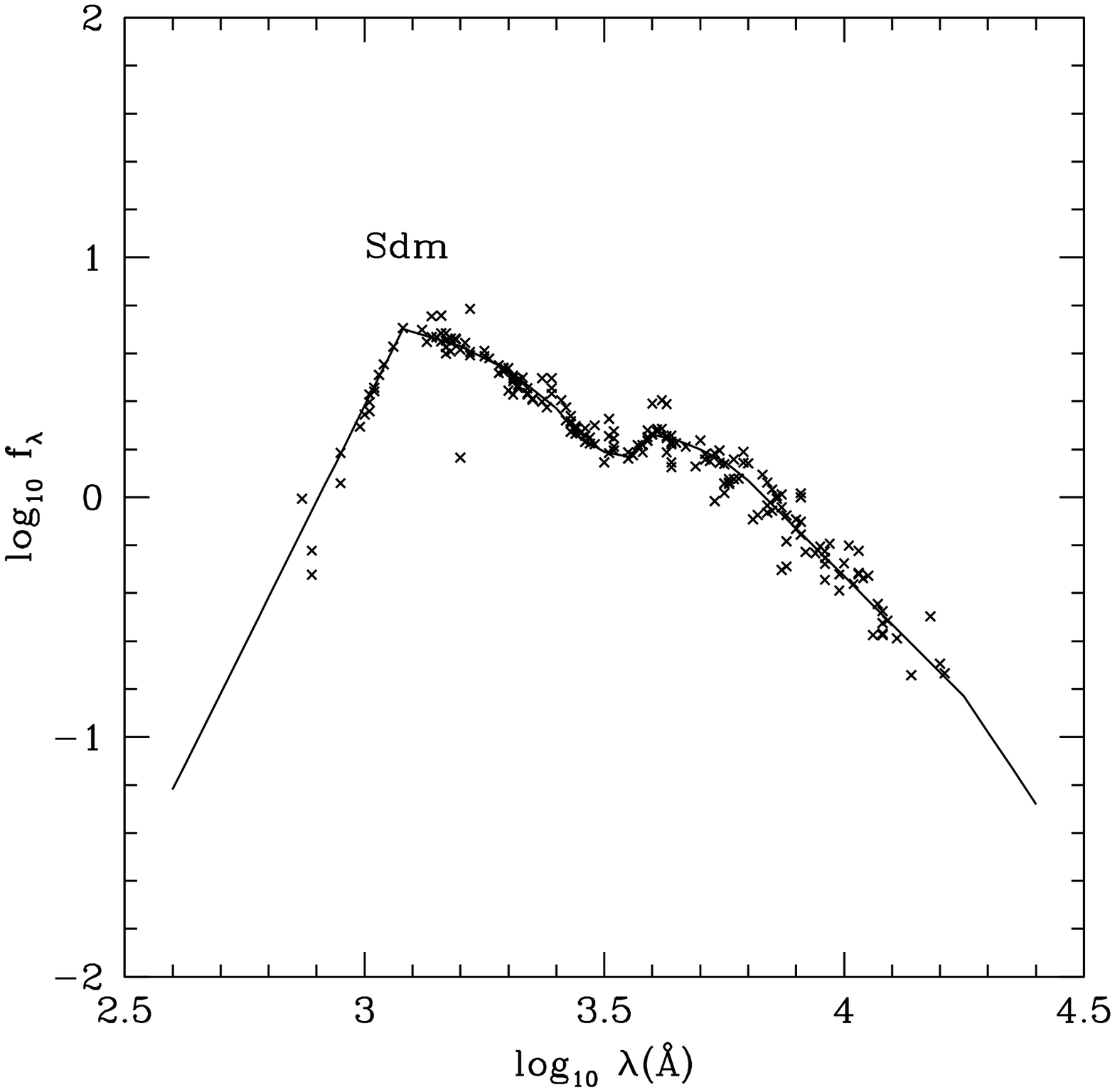,angle=0,width=8cm}
\caption{Assumed sed for Sdm galaxies, with data for galaxies with spectroscopic redshifts.}
\end{figure}

\begin{figure}
\epsfig{file=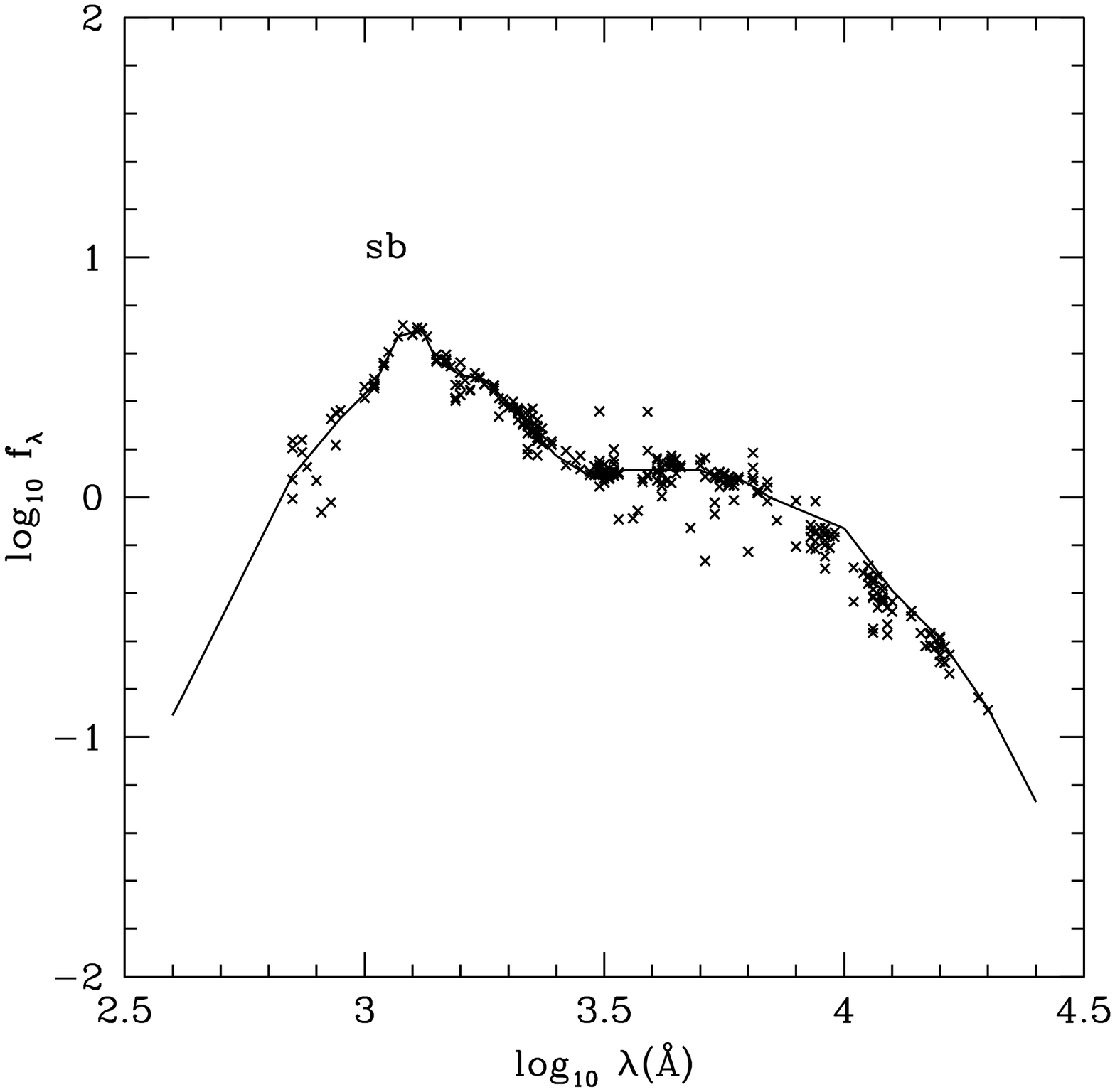,angle=0,width=8cm}
\caption{Assumed sed for starburst galaxies, with data for galaxies with spectroscopic redshifts.}
\end{figure}

\subsection{Photometric redshifts using the template method}
To try to extend the results of Connolly et al to a wider range of redshift, and to test
the robustness of the conclusions of Madau et al (1996), I have used the
photometric redshift method of Mobasher et al (1996) to analyze a much larger
sample of HDF galaxies, namely all those galaxies with $I_{AB} \leq$ 29.0 from
the catalogue of Williams et al (1996) detected
in at least the I and V bands, a total of 2490 galaxies once duplicates have
been eliminated.  I have also analyzed in detail the problem of aliasing, which is
central to the validity of the photometric redshift method.

As in Mobasher et al (1996) I use 6 galaxy spectral energy distributions, corresponding to
E, Sab, Sbc, Scd, Sdm and starburst galaxies, based on the seds of Yoshii and Takahara (1988)
and, for the starburst sed, Calzetti and Kinney (1992).  
The main changes from the work described in Mobasher et al (1996) are: 
(1) small modifications to seds, tuned to a preliminary list of 79 
spectroscopic redshifts, (2) inclusion of J, H, K data summarized by Fernando-Soto et al (1999),
 where available, (3) the solution is weighted with flux errors, where available 
 (but to avoid excessively high weighting by very high signal-to-noise observations, the
 minimum flux error is assumed to be 5$\%$ of the flux), 
(4) the permitted maximum redshift is increased to 6, (5) the range of M(B) is restricted to the
range -13.0 to -22.5 (to avoid excessive numbers of aliases at z = 0 and 6) , (6) dropouts are treated as follows:
if there is an upper limit in the U or B band which lies more than a factor 
4 below the flux (in nJy) in the next shortest wavelength band, this upper limit is used in the
photometric solution.


\begin{figure}
\epsfig{file=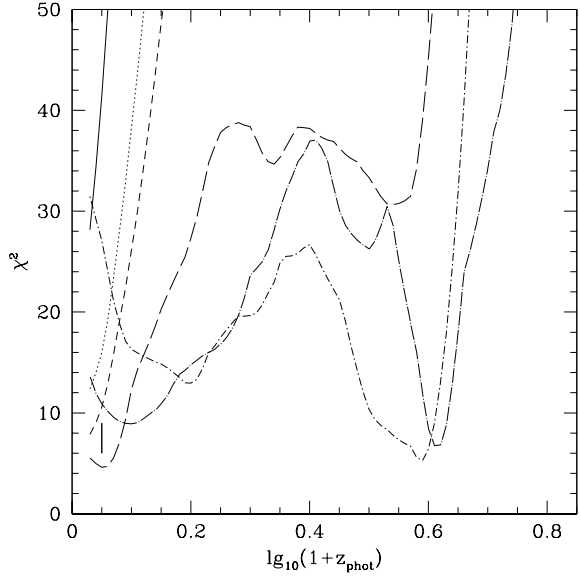,angle=0,width=8cm}
\caption{ Reduced $\chi^2$ (= $\chi^2/(n-2)$) as a function of $lg_{10}(1+z_{phot})$ 
for the 6 sed types,
for HDF 4-916.0 (n=7).  The vertical bar denotes the minimum $\chi^2$ at 
$lg_{10}(1+z_{phot})$ = 0.05, and there are aliases at 0.10, 0.20, 0.59 and 0.61,
of which the one at 0.59 is statistically significant.  There is no minimum corresponding to the reported spectroscopic
redshift at $lg_{10}(1+z_{spectr})$ = 0.280.}
\end{figure}


\begin{figure}
\epsfig{file=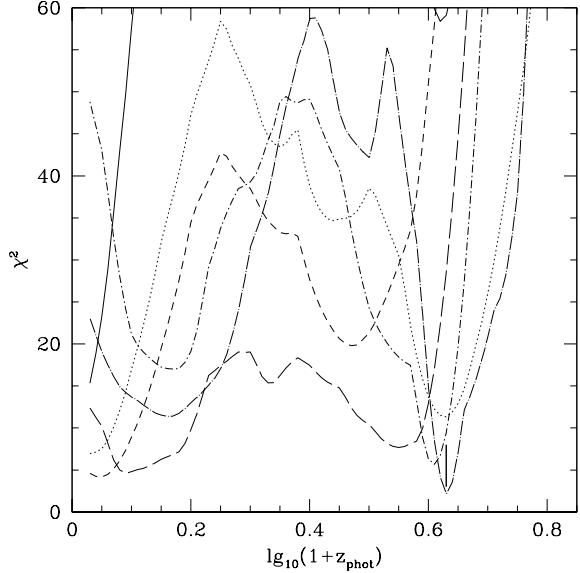,angle=0,width=8cm}
\caption{Reduced $\chi^2$ as a function of $lg_{10}(1+z_{phot})$ for the 6 sed types,
for HDF 4-52.111 (n=4).  The vertical bar denotes the minimum $\chi^2$ at 
$lg_{10}(1+z_{phot})$ = 0.63, and there are aliases at 0.05, 0.09, 0.55 and 0.61, none of 
which are statistically significant. The spectroscopic
redshift is $lg_{10}(1+z_{spectr})$ = 0.595.  The FLY value is 0.107.}
\end{figure}


\begin{figure}
\epsfig{file=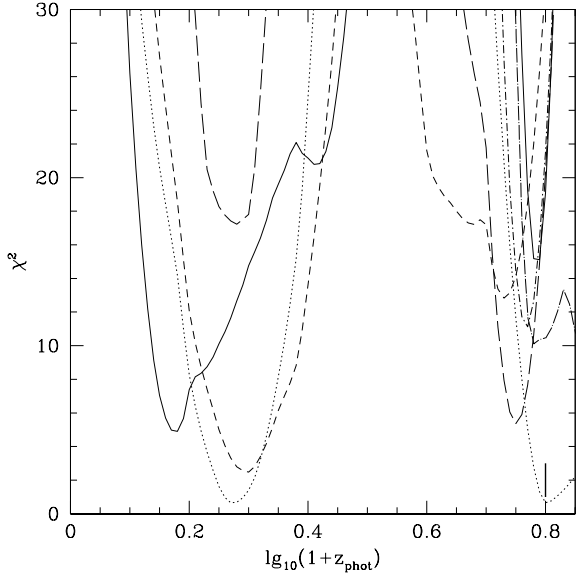,angle=0,width=8cm}
\caption{Reduced $\chi^2$ as a function of $lg_{10}(1+z_{phot})$ for the 6 sed types,
for HDF 3-387.0 (n=4).  The vertical bar denotes the minimum $\chi^2$ at 
$lg_{10}(1+z_{phot})$ = 0.80, and there is a significant alias at 0.27.  The FLY value is 0.265.}
\end{figure}

\begin{figure}
\epsfig{file=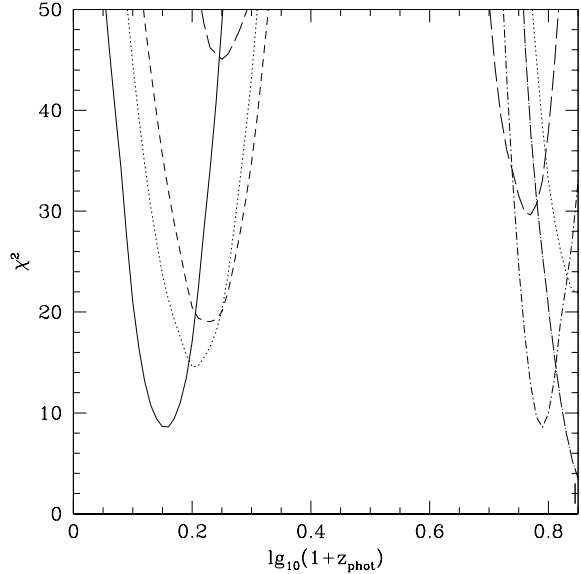,angle=0,width=8cm}
\caption{Reduced $\chi^2$ as a function of $lg_{10}(1+z_{phot})$ for the 6 sed types,
for HDF 1-71.0 (n=6).  The vertical bar denotes the minimum $\chi^2$ at 
$lg_{10}(1+z_{phot})$ = 0.85, and there are aliases at 0.16 and 0.76,
but not statistically significant.  The FLY value is 0.739.}
\end{figure}

\begin{figure}
\epsfig{file=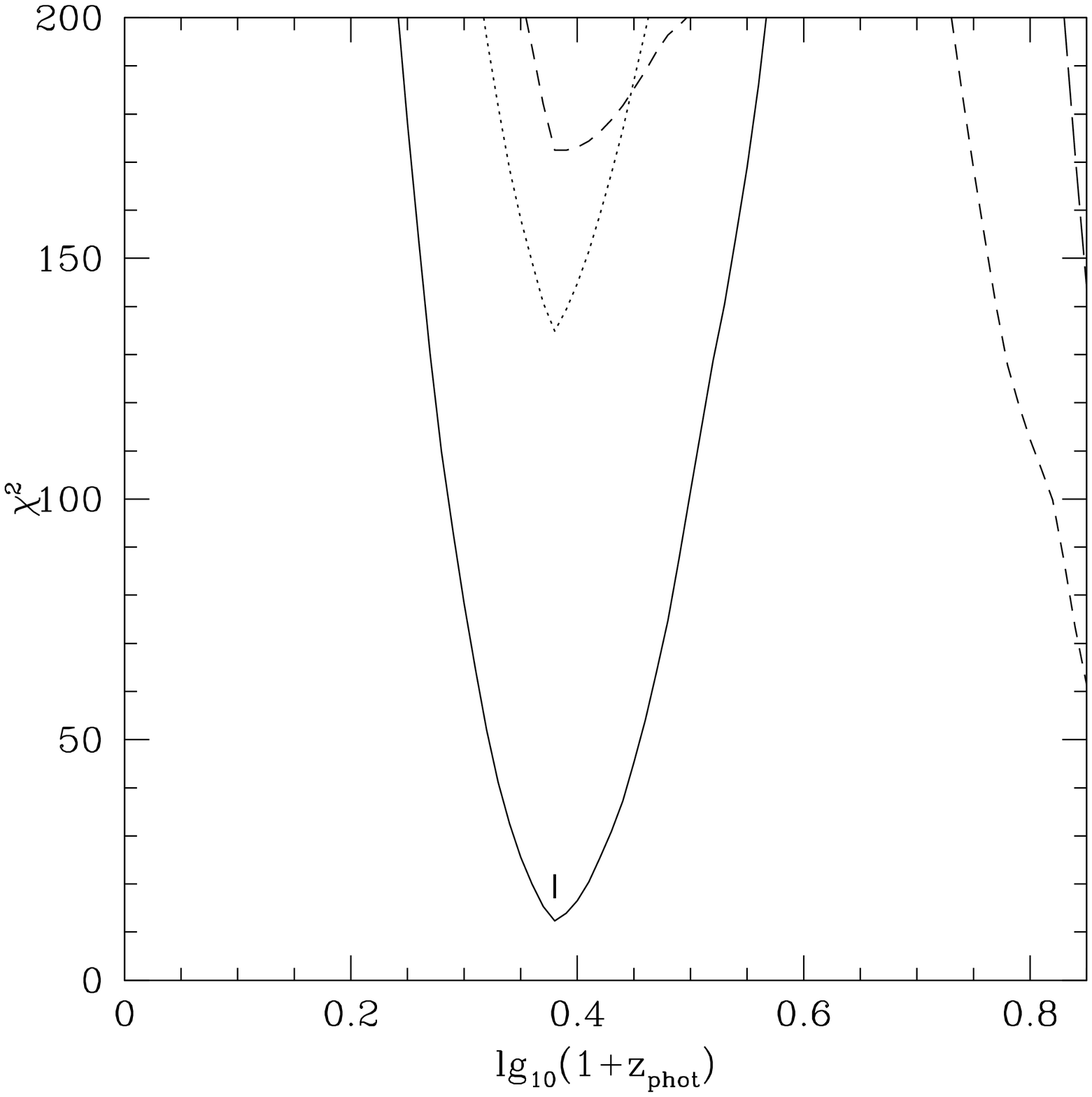,angle=0,width=8cm}
\caption{Reduced $\chi^2$ as a function of $lg_{10}(1+z_{phot})$ for the 6 sed types,
for HDF 4-403.0 (n=6), the galaxy in which a high redshift supernova has been observed (Riess et al 2001).  
The vertical bar denotes the minimum $\chi^2$ at 
$lg_{10}(1+z_{phot})$ = 0.38 (z = 1.4 $\pm$0.05).  This is consistent with the estimate by Budivari et al (2000) of
z = 1.55 $\pm$ 0.15 for this galaxy, but only marginally with the estimate 1.7 $\pm$0.1
given for the supernova by Riess et al (2001).}
\end{figure}

The seds used are shown in Figs 1-6, compared with data for galaxies with known spectroscopic 
redshifts.  Because different parts of these seds are sampled by the HDF UBVI bands at different 
redshifts, different frequency ranges of these seds have validity over different redshift ranges.
The number of seds used is quite critical to the success of the solution.  Too few leads to 
what Bolzonella et al (2000) term 'catastrophic' failures, too many leads to increased aliasing
and hence again to failures.  
Figs 1-6 illustrate that the 6 sed templates give a good low-resolution representation of the 
observed seds.  Although a detailed treatment of the evolution of galaxy seds would 
improve the spectral resolution of the seds and would probably improve the detailed fits,
there is also likely to be an increased incidence of aliasing unless the observational 
data is also at improved spectral resolution.  This seems to be borne out be the studies
of Bolzanella et al (2000) and Le Borgne and Rocca-Volmerange (2003) (see section 2.3).

The percentage of galaxy sed types as a function of redshift, for $z < 2.5$, is shown in Table 1.
The percentage of ellipticals drops at $z > 1$, as found by Rodighiero et al (2001) from
imaging studies.  The fraction of starburst galaxies 
increases steadily from 23$\%$ at z$<$0.3 to $>$50$\%$ at z$>$1, consistent with the observed
morphologies found by Lilley et al (1996).  At higher redshifts the sed type is determined 
mainly by the sed of the most recent starburst and so has less connection with Hubble types.

One of the main problems in photometric redshift methods is aliasing. For a given set of 
photometric data, there may be two or more distinct redshifts, often, but not always, from
different sed templates, which give an almost equally good representation of the data.
Increasing the number of galaxy templates, and introducing other free parameters in the solution
like extinction, may improve the accuracy of the redshift determination for the majority of galaxies,
but at the expense of increasing the number of aliases, and hence of completely wrong redshifts. 
To try to make a quantitative estimate of the probablity of aliasing, I proceed as follows.
For each galaxy type and redshift bin, in the range 0.03 (0.01) 0.85 for $lg_{10} (1+z_{phot})$,
I calculate $\chi^2$, and look for the minimum.  I define a significant alias as a value of
$lg_{10} (1+z_{phot})$ differing from the minimum value by more than 0.08, ie (1+z) differs by more
than 20 $\%$, with $\chi^2$
no more than 1.0 above the minimum value (see figs 7-11 for examples).  
The proportions of galaxies with significant aliases
then depends very strongly on the number of photometric bands in the solution, n, where presence
of a significant dropout is included as a detected band.  For
n = 2, 3, 4, 5, 6, 7, the proportions are 99, 38, 1.5, 0.5, 0.0, 1.5 $\%$.  Thus a minimum of 4 bands
is needed for effective estimation of photometric redshifts.  Very little weight can be attached to
estimates based on 2 bands, of which there are 374 in our sample of 2470.

\begin{figure}
\epsfig{file=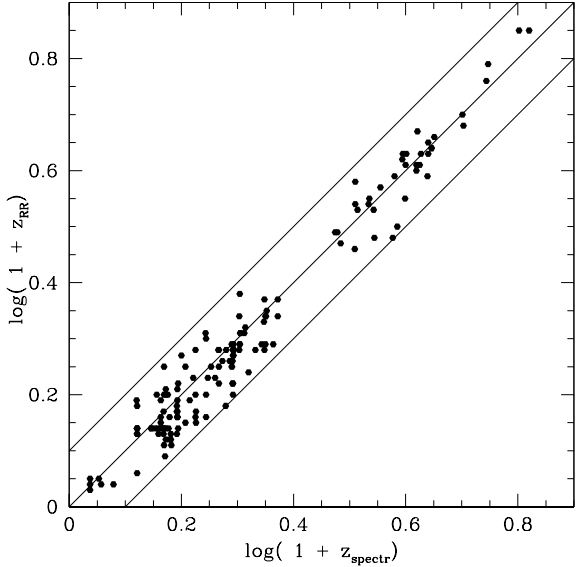,angle=0,width=8cm}
\caption{Photometric versus spectroscopic redshifts for 152 galaxies in the
Hubble Deep Field (N), from template method of present paper.
The rms value of $log_{10} [(1+z_{ph})/(1+z_{sp})] $ is 0.0398 , corresponding to a
9.6 $\%$ accuracy in photometric estimates of (1+z).}
\end{figure}

\begin{figure}
\epsfig{file=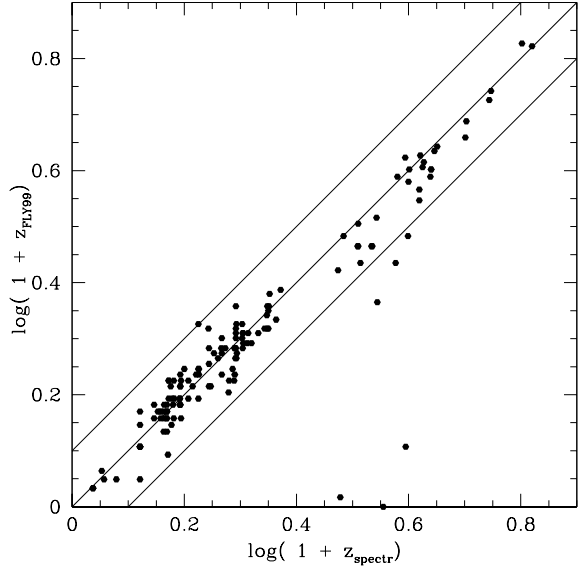,angle=0,width=8cm}
\caption{Photometric versus spectroscopic redshifts for 145 galaxies in the
HDF-N, from Fernandez-Soto et al (1999).}
\end{figure}

\begin{figure}
\epsfig{file=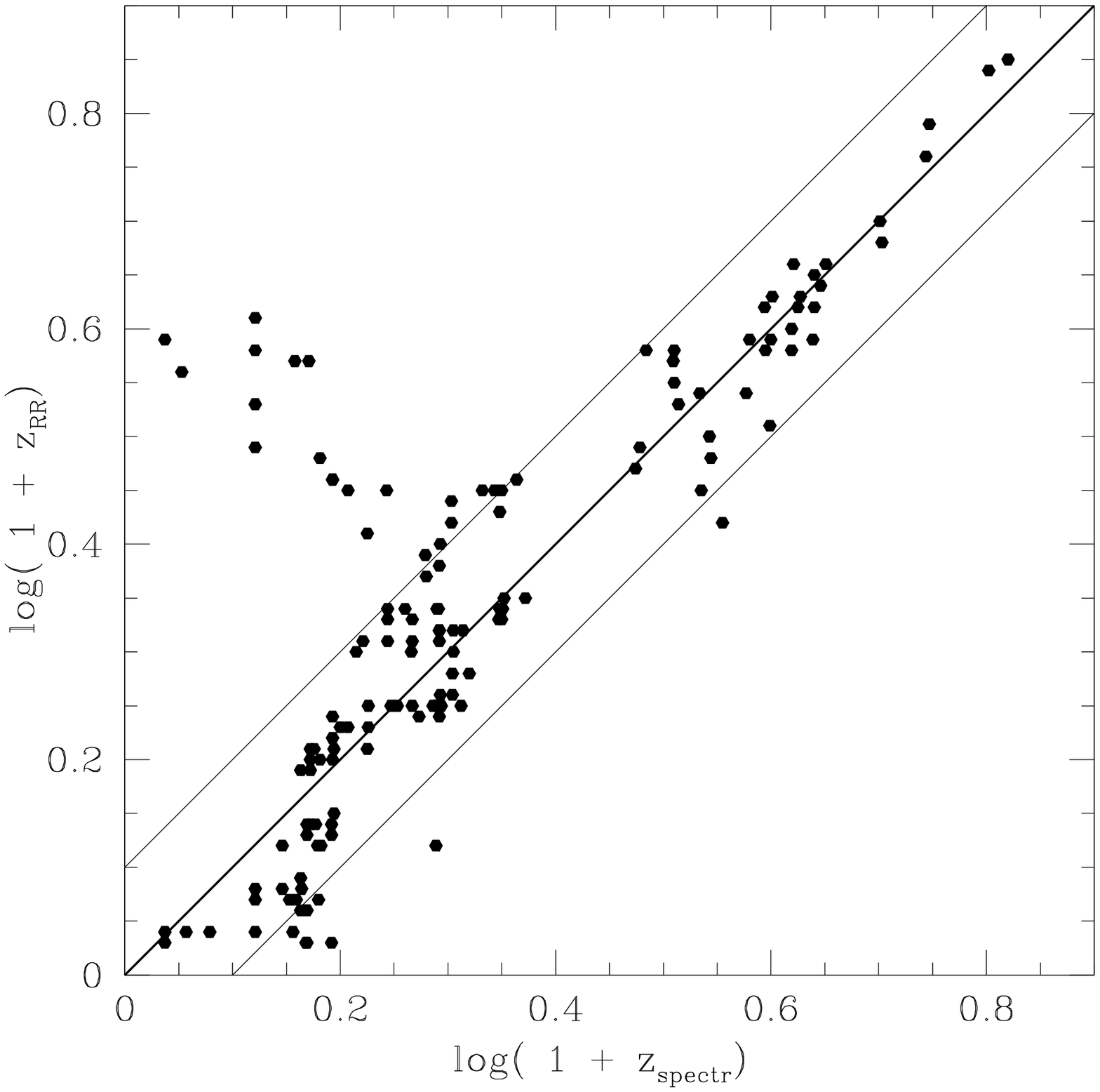,angle=0,width=8cm}
\caption{As for Fig 12, but using only 4 bands (UBVI).}
\end{figure}

\begin{figure}
\epsfig{file=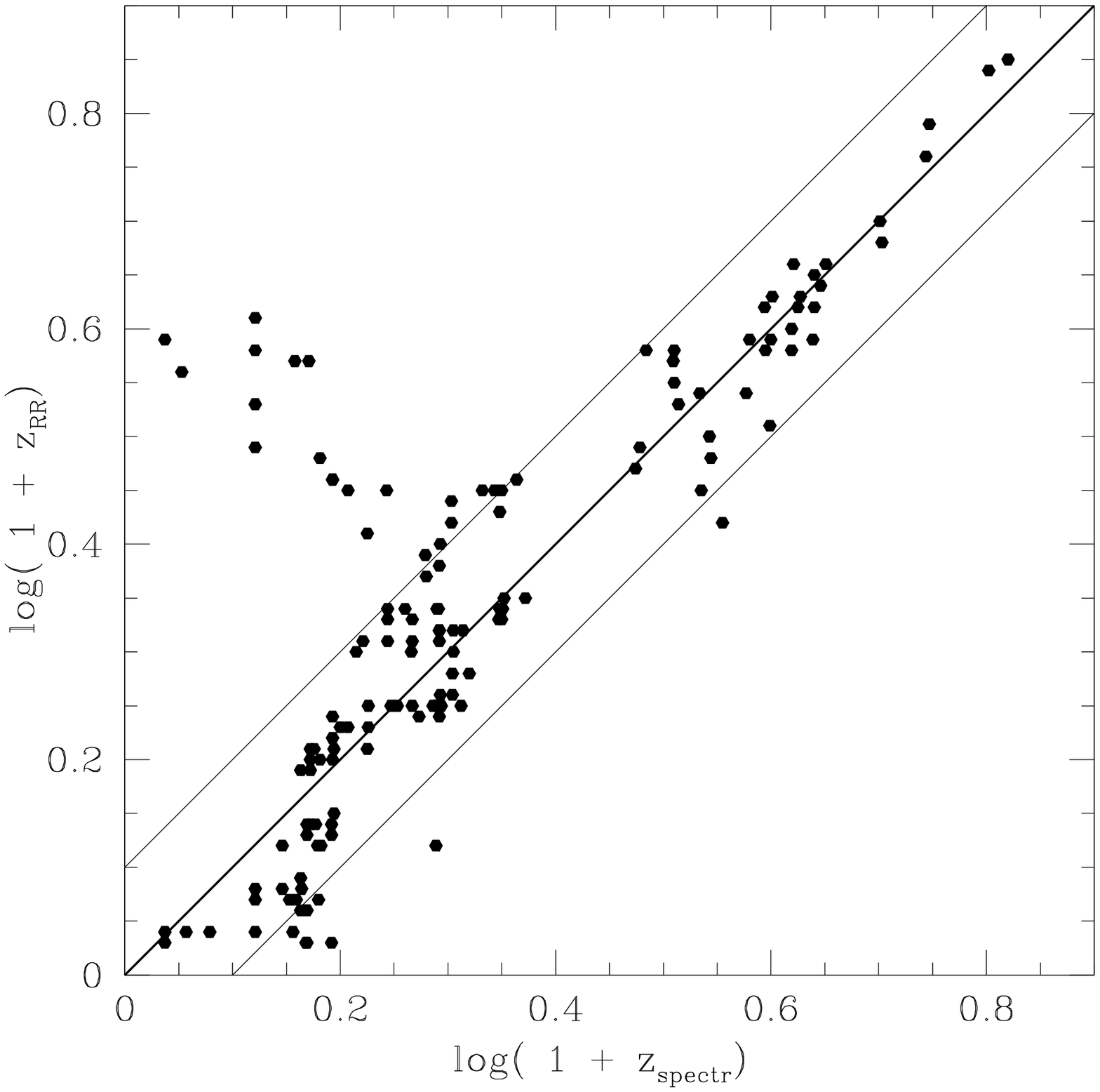,angle=0,width=8cm}
\caption{As for Fig 12, but using only 3 bands (UBV).}
\end{figure}

\begin{figure}
\epsfig{file=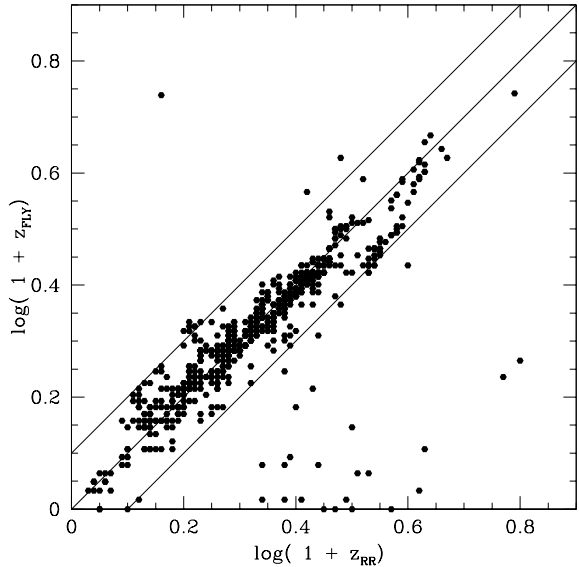,angle=0,width=8cm}
\caption{Comparison of photometric redshifts from present paper with those of FLY,
for sources detected in at least 4 bands.}
\end{figure}

\begin{figure}
\epsfig{file=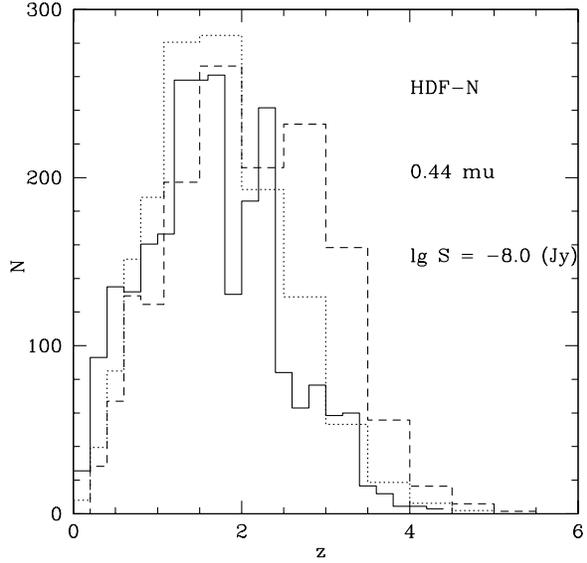,angle=0,width=8cm}
\caption{Redshift distribution for the Hubble Deep Field (N).  The solid histogram is
derived from photometric redshifts, as described in section 3, for B $<$ 29.  The broken histogram is
the prediction from the $\lambda_o$ = 0.7 model of Rowan-Robinson (2001), for S(0.44 $\mu$m) = 10 nJy
and the dotted curve is the corresponding model of King and Rowan-Robinson (2003).  }
\end{figure}

\begin{figure}
\epsfig{file=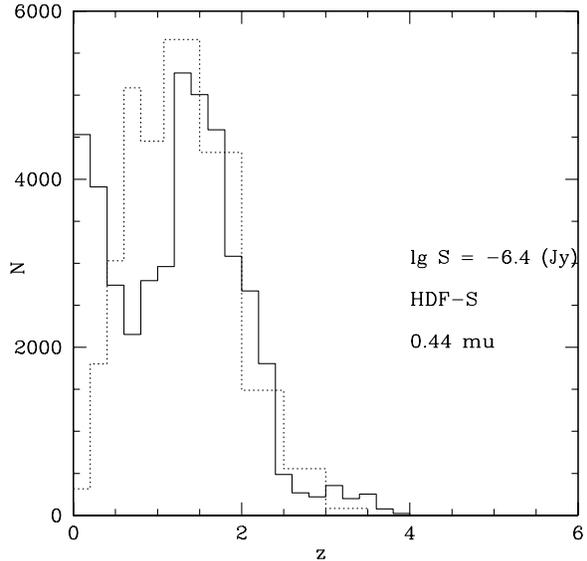,angle=0,width=8cm}
\caption{Redshift distribution for the Hubble Deep Field (S).  The solid histogram is
derived from photometric redshifts, as described in section 3, for B $<$ 25.  The broken histogram is
the prediction from the $\lambda_o$ = 0.7 model of Rowan-Robinson (2001), for S(0.44 $\mu$m) = 400 nJy.  }
\end{figure}

The comparison of photometric and spectroscopic redshifts for
152 HDF galaxies with spectroscopic redshifts in the literature is shown in Fig 12.
The sample consists of 145 redshifts given in the data set of 1067  HDF galaxies given by Fernandez-Soto et al
(1999, hereafter FLY), with the deletion of 4-555.2 and 3-550.1 (no longer listed as spectroscopic
redshifts in Fernandez-Soto et al 2001 or Cohen et al 2000), the addition of 4-618.0 (Cohen 2001),
and the corrected redshifts for 4-852.12, 2-906.0 and 2-256.0 given by Cohen (2001).
In addition spectroscopic redshifts for 5 galaxies, for which all photometric redshift methods agree
on a different redshift and find no alias at the quoted spectroscopic redshift, are deleted:
4-948.11111, 4-878.11, 3-355.0, 4-316.0, and 4-916.0. For the remaining 1423 galaxies in
the Williams et al (1966) catalogue satisfying the constraints given at the beginning of section 2.1, 
there are a further 7 spectroscopic redshifts available. 

My template method gives redshifts accurate to 9.6 $\%$ in (1+z).  Bolzonella et al (2000)
report an increase in the uncertainty in $z_{phot}$ with redshift from their simulations
but this does not seem to be the case for the more meaningful quantity
$(z_{phot}-z_{spect})(1+z_{spect})$.  

Some individual discrepancies between 
photometric and spectroscopic redshifts are summarized in Table 2.  For 7 objects
my estimate agrees with the spectroscopic redshift within 0.1 in $log_{10}(1+z)$,
but the Fernando-Soto et al (1999) estimate disagrees: 4 of these have been fixed by the
revisions applied in Fernandez-Soto et al (2001).  On this basis my photometric
redshift estimates appear marginally better than those of Fernando-Soto et al.
The concordant redshifts in Fig 12 include 7 cases where only 3 bands
were detected and 2 where only 2 bands were detected, so even though these galaxies have
significant aliases the method has selected the correct redshift. Figure 13 shows the corresponding
comparison for Fernandez-Soto et al (1999).  Figure 14 shows what happens to my photometric
redshift estimates if I use only the 4 bands U, B, V, I.  For only one galaxy does the 
photometric redshift become strongly discrepant, confirming that for a high proportion of cases
4-band estimates are reliable.  On the other hand if only 3 bands (U, B, V) are used (Fig 15), 
about 20$\%$ of the photometric estimates are discrepant.  This is entirely consistent with
the success that Steidel et al (1998) have had using the Lyman dropout technique with
3 photometric bands, with 75 $\%$ of their U-band dropout galaxies have spectroscopic
redshifts in the range 2.7-3.3.  

Fig 16 shows a direct comparison of photometric redshift estimates by Fernandez-Soto et al (1999) and myself, for 
galaxies detected in at least four photometric bands.  The agreement is extremely good, with only 38/688
estimates being discrepant.  The group
of galaxies for which Fernandez-Soto et al give significantly lower redshift estimates 
include three galaxies in Table 1 for which the spectroscopic redshift appears to confirm
my estimate.  The greater scatter appearing in the comparison of photometric redshift estimates
by Fernandez-Soto et al and myself made by Peacock et al (2000, their Fig 1) arises because this
comparison includes estimates based on only 2 or 3 photometric bands.  Table 3 lists the 33 galaxies
which I estimate to have z $>$ 5, together with spectroscopic measurements (3 cases) and estimates
by Fernandez-Soto et al (1999).  For 4 of the galaxies Fernandez-Soto et al estimate low redshifts:
these are all cases where I find a significant alias (an example is shown in Fig 9).  Otherwise
the agreement is reasonable, but with a tendency for my redshifts to be slightly higher than those of
Fernandez-Soto et al (and than the spectroscopic redshifts, where available).

Fontana et al (2000) find a rather poor performance of their photometric redshift estimates when only 
four photometric bands are used (their Fig 10), in contrast to the excellent results obtained here.
Similarly the Bolzonella et al (2000) 'hyperz' code has quite a significant proportion of failures with
4 or even 5 bands (their Fig 2).  I also have no problems with the 2 galaxies 4-473.0 and 4-52.111 
(see fig 8, present paper), for which
Bolzonella et al (2000) report catastrophic failure with 6 different sed prescriptions
(their Fig 4).  I conclude that my code is more robust than those of Fontana et al (2000) and 
Bolzonella et al (2000).

My photometric redshift estimate for
HDF 4-403.0, in which Riess et al (2001) have detected a supernova, is 1.4 $\pm$0.05 (see fig 11),
consistent with the estimate by Budavari et al (2000) of 1.55 $\pm$0.15 but only marginally consistent
with the estimate of 1.7 $\pm$0.1 derived by Riess et al (2001) for the supernova.

The redshift distribution derived above for HDF-N galaxies with B $<$ 29
is shown in Fig 17, compared with the predictions of the models of Rowan-Robinson (2001) and 
King and Rowan-Robinson (2003) for an $\lambda_o$ = 0.7 universe.    
The King and Rowan-Robinson (2003) model fits well, while the Rowan-Robinson (2001)
model predicts too many galaxies at z $>$ 2.5.

\begin{figure*}
\epsfig{file=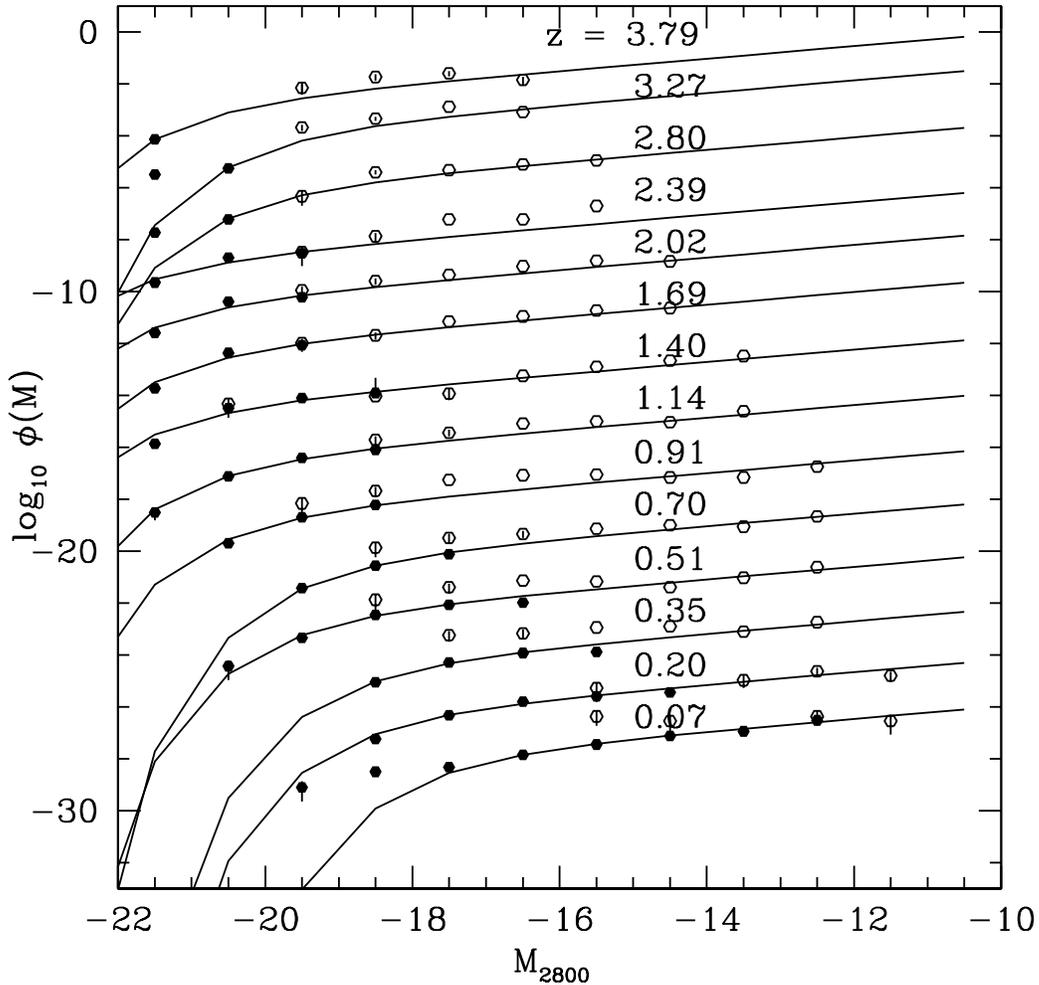,angle=0,width=14cm}
\caption{2800 $\AA$ luminosity functions derived from HDF-N (open circles) and HDF-S (filled circles)
photometric redshifts, in bins of 0.05 in $log_{10}(1+z)$.
No correction for dust extinction has been applied at this stage.
The solid curves are best-fitting Schecter functions, with a faint-end slope $\alpha$ = 1.6.
An $\Omega$ = 1 universe is assumed.}
\end{figure*}

The Goddard Space Flight Center Group (Teplitz et al 2001) have published UBVRI data for galaxies
in a 0.5 square degree area centred on HDF-S.  After elimination of objects characterized as stars,
there are 28719 galaxies in their $5-\sigma$ catalogue and I have used these to estimate
photometric redshifts as above.  I have 
compared my estimates with the spectroscopic redshifts of Glazebrook et al (2001), and like 
Teplitz et al (2001) I find
good agreement if the comparison is restricted to Glazebrook et al's quality 3 and 4 redshifts.
Figure 18 shows the photometric redshift distribution for galaxies
from this catalogue with B $<$ 25, which appears to be the completeness limit, compared with
predictions of the model by Rowan-Robinson (2001).  The agreement is good except at low redshifts,
where the star-formation history derived for these data differs significantly from that assumed in
the model (see Fig 28): this merits further investigation. 

The combined 2800 $\AA$ luminosity function for HDF-N and HDF-S is then estimated for each redshift bin and 
fitted with a Schecter function.  No correction for the effects of dust extinction has been applied at this stage.
The results are summarized in Fig 19   for an $\Omega_o$ = 1 universe.  The uncertainties 
given combine in quadrature the purely statistical uncertainties and the additional
contribution to the scatter due to the use of photometric redshifts, which has been
estimated from a series of simulations following the method of SubbaRao et al (1996).
The distribution of $log_{10}(1+z)$ was assumed to be gaussian, with $\sigma$ = 0.04
where 4 or more bands are available, 0.06 if only 3 bands are available, and 0.08
if only 2 bands are available, to take account of the increased uncertainty where less than 4 bands
are available.  There are some discrepancies between estimates from HDF-N and HDF-S at
luminosities where they overlap, but overall the agreement of the two datasets with each other, and with
the models, is good.

I find that the best faint end slope of the luminosity function for the redshift range 0-4 is $\alpha$ = 1.6, 
which is also the value found by Steidel et al (1999) for galaxies with z = 3.  
Figure 19 shows results assuming $\alpha$ = 1.6.
Most local studies at optical wavelengths find faint end slopes in the range $\alpha$ = 1.1-1.3. 

Lanzetta et al (2002) have used the HDF-N sample to study the surface brightness distribution function.
They note that uncertainty in the shape of the faint end of this distribution at higher redshifts translates into 
a major uncertainty in the star-formation rate.  In the present work, the larger HDF-N sample and the
use of both HDF-N and -S samples provides sufficient dynamic range in luminosity, even at high redshift, 
that the uncertainty in the shape of the luminosity function and in the star-formation rate is less severe.

My values for $M_{*,2800}$  at z =  0.70, 1.14, 1.40, 1.69 (-18.5, -20.0, -20.6, -20.4) are reasonably
consistent with those found by Connolly et al (1997) for z = 0.5-1, 1-1.5, 1.5-2 (-18.75, -19.5, -20.25).
In section 3, I will use these luminosity functions to estimate the evolution of the star-formation
rate in galaxies, but first we need to consider the evolution of extinction in galaxies.

\subsection{Effect of inclusion of $A_V$ as a free parameter}

The semi-empirical sed templates used above can be assumed to already take account of an
average amount of internal extinction at z = 0.  However we need to allow for
the possibility of evolution of the average extinction with redshift, an
effect predicted by Pei et al (1999) and Calzetti and Heckman (1999) for
a variety of star formation histories.  It is also clear from far infrared
and submillimetre surveys that some galaxies have significantly higher
extinctions than the average for their Hubble type (see also the study of
local galaxies by Rowan-Robinson 2003).

Le Borgne and Rocca-Volmerange (2002) include in their models allowance for the
evolution of the characteristic dust extinction with time.  However to fully allow
for the variation in dust extinction from galaxy to galaxy we really need to 
solve for $A_V$ as a free parameter, in addition to the redshift.  Since we are trying
to do this with rather limited photometric data, this will inevitably result in
increased aliasing.  Basically an earlier Hubble type may look similar, over a
restricted wavelength range, to a later type with substantial reddening.  This
aliasing problem seems to be born out by the study of Bolzanella et al (2001),
who include $A_V$ as a free parameter and have significantly worse aliasing
problems than the pure template solution of section 2.1 above.

To try to control this aliasing problem to some extent, I have made the following
restrictions: (i) I assume that there is no extinction in ellipticals. 
(ii) $A_V$ is allowed to vary between -0.4 and + 1.0.  Negative values
are needed because we may expect $A_V$ to be lower at high redshift than at the 
present epoch.  The assumption is that the typical value of $A_V$ at the
present epoch in spirals is 0.3-0.4 mag (Rowan-Robinson 2003).  The $A_V$ we are solving for is the
difference between the actual value and the mean value at the present epoch.
(iii) No solution for $A_V$ is sought if
the reduced $\chi^2$ for the solution with $A_V$ = 0 is $<$ 0.01 or if there are 
less than 3 bands available.  (iv) A non-zero $A_V$ is accepted only if the
reduced $\chi^2$, allowing for the reduced number of degrees of freedom, is
improved by the inclusion of $A_V$. (v)  A prior expectation that the 
probability of a given value of $A_V$ declines as $|A_V|$ moves away from
zero is introduced by minimizing  $\chi^2$ + $\alpha A_V^2$ rather than
$\chi^2$.

\begin{figure}
\epsfig{file=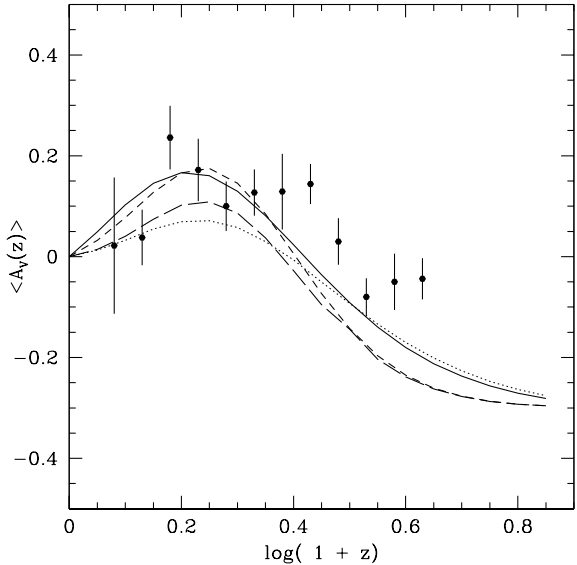,angle=0,width=8cm}
\caption{Variation of mean value of $A_V$ with redshift, derived from HDFN.  
Curves are theoretical models calculated as described in Appendix, assuming that $A_V(0) = 0.3$
(Rowan-Robinson 2003).  Solid curve: $\Omega_0 = 1$ (dotted curve: with density evolution as in
King and Rowan-Robinson 2003); broken curves: $\lambda_0 = 0.7$.}
\end{figure}

Fig 20 shows the variation of $<A_V(z)>$ with redshift, 
calculated from different galaxy samples,
compared with the prediction of closed box star formation models of the form used by Rowan-Robinson (2001)
and King and Rowan-Robinson (2003) (see Appendix).  There is some
evidence for the expected increase in $<A_V(z)>$ towards z = 1, and
of the expected decrease at higher redshift.  Fig 21 shows the distribution of $\chi^2$, with and without the inclusion
of $A_V$.  The inclusion of $A_V$ does reduce many of the higher values of $\chi^2$.

The inference that the extinction in z $\sim$ 3 galaxies is smaller than in galaxies locally is at odds with the
claim of Steidel et al (1999), supported by Adelberger and Steidel (2000) and Vijh et al (2003), that the
extinction at z $\sim$ 3 is substantial, E(B-V) = 0.15.  However it is noteworthy that these latter claims are based on a
single colour, (G-R).  In the case of the HDF-N sample we can do better than this, because most of the
z $\sim$ 3 galaxies are detected at 4500, 6000 and 8000 $\AA$.  Fig 22 shows the [6000-8000] v. [4500-6000] 
colour-colour diagram for HDF-N galaxies with z = 3 $\pm$ 0.4.  The locations of my six galaxy templates
corresponding to z = 3 are indicated, as also is a reddening line from the starburst locus (sb).  While
some of the spread can be attributed to reddening, with E(B-V) $\leq$ 0.1, the main elongation of the
distribution is consistent with being due to different recent star-formation histories.  Note the the
location of the elliptical component (E) is due to the effect of planetary nebulae, not recent star-formation.
Vijh et al (2003) claim that different star-formation histories can not account for the broad spread in (G-R)
colours seen in the Lyman drop-out galaxies, but the models they consider allow only for different times
since the start of star-formation.  If we look instead at models for an instantaneous burst viewed at
different times from the end of the burst (Bruzual and Charlot 1993, Fig 4a, or Bruzual 2000, Fig 1a), it is 
easy to see that the (G-R) colour from a z = 3 galaxy changes by at least 2 magnitudes in 0.3 Gyr.  However
the age-dust degeneracy can be broken if infrared observations are available and Shapley et al (2001) do
find consistency with the Steidel et al (1999) E(B-V) estimates from a sub-sample of 63 z $\sim$ 3 galaxies
with J and K measurements.  Erb et al (2003) find better agreement between star-formation rates derived
from uv continuum and H$\alpha$ for 16 galaxies at 2.0 $<$ z $<$ 2.6 if a mean extinction of E(B-V) = 0.10
is corrected for.

\begin{figure}
\epsfig{file=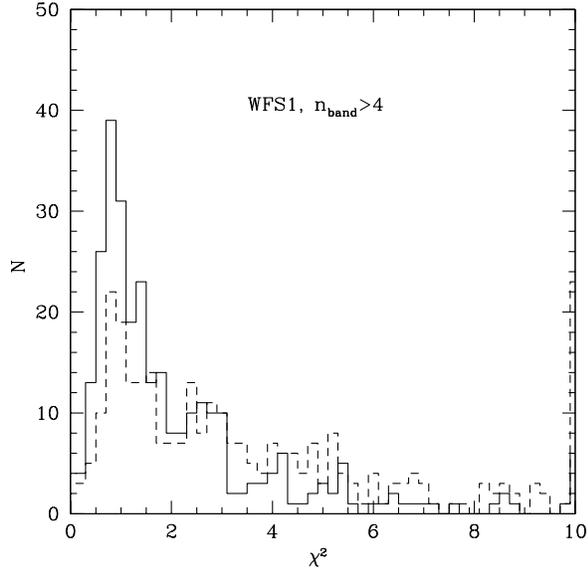,angle=0,width=8cm}
\caption{Plot of distribution of $\chi^2$ with no correction for $A_V$ (broken line), and including
effect of $A_V$ (solid line).}
\end{figure}

\begin{figure}
\epsfig{file=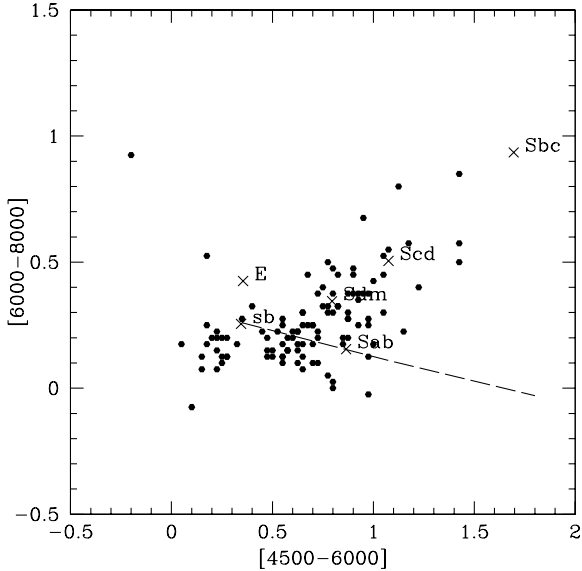,angle=0,width=8cm}
\caption{[6000]-[8000] versus [4500]-[6000] for HDF-N galaxies with z = 3.0 $\pm$ 0.4.
Loci for template  galaxies at z = 3, together with reddening line for starburst template (extending to E(B-V) = 1.0)
are shown.}
\end{figure}

My photometric redshift catalogues for HDF-N, HFF and HDF-S, with parameters estimated by the
template method both with $A_V$ set = 0, and with $A_V$ as a free parameter, are available at

http://astro.ic.ac.uk/$\sim$mrr/photz .

\subsection{Inclusion of treatment of sed evolution}

Mobasher and Mazzei (1998) and Le Borgne and Rocca-Volmerange (2002) have used spectral synthesis
models to track the evolution of seds with redshift.  In principle this should
yield improved photometric redshifts compared with a fixed template method, since the
latter can only be accurate over a fixed range of photometric bands and a narrow range
of redshift.  Fig 23 shows the excellent results reported by Le Borgne and Rocca-Volmerange (2002)
for HDF-N galaxies with photometric data in 7 bands.  However the inclusion of
additional parameters in the modelling to characterize the star-formation history inevitably
results in increased aliasing as can be seen in Fig 23 and Table 2.

\begin{figure}
\epsfig{file=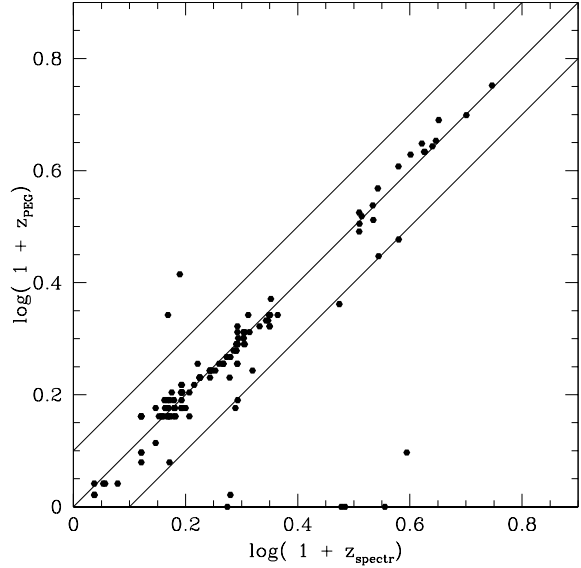,angle=0,width=8cm}
\caption{$z_{phot}$ versus $z_{spect}$, from Le Borgne and Rocca (2002).}
\end{figure}

It is also of interest to ask whether galaxies do really represent a small set of star-formation
histories, as assumed by Le Borgne and Rocca-Volmerange (2002).  Their main parameter 
characterizing seds is $\nu$, where SFR = $\nu M_{gas}$ (they also have assumptions
about infall and galactic winds).  Clearly $\nu$ determines the ratio 
b = ${M}_*/\dot{M}_* t_0$
which was shown by Larsen and Tinsley (1974) and Scalo (1986) to characterise the Hubble sequence.
In practice, as can be seen from Fig 4 of Bruzual
and Charlot (1993), galaxy seds consist of two almost independent components; (a) stars
formed more than 1 billion years ago, which for a fixed IMF produce an sed peaking in
the near infrared which is almost independent of the history of how those stars were
formed and only carries (at low resolution at any rate) information on the total mass
of stars formed, (b) stars formed less than a billion years ago, which generate an
sed dominating the blue and uv part of the spectrum.  This part of the sed depends
sensitively on the details of the recent star formation rate and whether there have been
recent starbursts.  It seems possible that the form of the blue-uv part of the spectrum may 
be uncorrelated with b, since there is no guarantee that the past billion years of
a galaxy's life are typical of its long-term past.

I have therefore investigated deconvolution of galaxy seds into these two components (a) and
(b), extending an approach followed by Rowan-Robinson (2001) in modelling galaxy counts.
I assume that the elliptical galaxy sed represents component (a) and have then subtracted
a multiple of this from the other 5 sed types to fit their red-ir spectrum.  The 5 blue-uv residues
are taken to be representative of typical recent star formation histories.  Each galaxy sed 
is then fitted with an arbitrary combination of components (a) and (b), with the 
normalisation constants essentially specifying $M_*$ and $\dot{M}_*$, respectively.
However it is convenient to characterise the ratio of old to young stars by a parameter
$f_{old}$ such that this is 1 when the sed exactly reproduces the original sed template.
I have applied this method only to the HDF-N 7-band data because we need near infrared data to get 
a good determination of $M_*$.  This method should be very powerful when applied to SIRTF-IRAC 
data, especially where multiband optical data is also available.

A small number of additional aliases (see Table 2) are generated by the inclusion of this new
parameter $f_{old}$, which I allow to range logarithmicly between 0.04 and 25.

Figure 24 shows a plot of $M_*/\dot{M}_* t_o$ versus redshift, with the curves corresponding
to different star-formation scenarios, varying the value of the exponential parameter Q
used by Rowan-Robinson (2001) - see Appendix.  The broad trend is as expected, with higher values
of b towards the past.  This is a more powerful demonstration of the evolution of galaxy seds 
than merely showing that consistent redshifts are obtained if an evolutionary scenario is used.
A wide range of value of Q is needed to understand the distribution of b with z.  Star-formation 
was assumed to be initiated at $z_i$ = 10 for this family of curves.  To understand galaxies with very 
low values of b (very weak bulge components) at z = 0.5-2 it is necessary to assume that some galaxies
 started to form stars only at much later redshift.

\begin{figure}
\epsfig{file=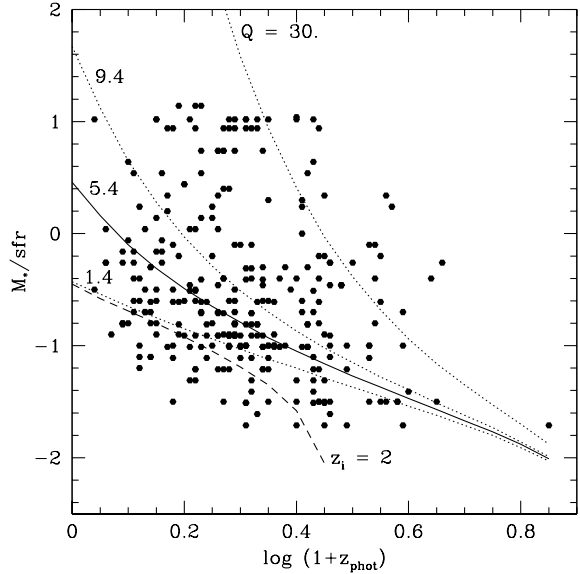,angle=0,width=8cm}
\caption{Plot of $M_*/\dot{M_*} t_0$ versus redshift.  Curves for simple star-formation histories
of the form used by Rowan-Robinson (2001) are shown for Q = 1.4, 5.4, 9.4 and 30, with P = 1.2,
$z_i$ = 10, $\Omega_o$ = 1, in each case.  The broken curve corresponds to Q = 1.4, $z_i$ = 2.}
\end{figure}

Figure 25 shows a plot of $M_*/\dot{M}_* t_0$ versus uv sed type, showing a general trend but 
poor correlation in detail.  If there were a true Hubble sequence characterised only by
$\nu$ or Q, then the points would have fallen on a single locus.  In fact there is a great deal 
of scatter and a wide range of uv sed types can correspond to a single value of b.  
Thus there is not a very strong correlation between the recent star-formation
history of a galaxy and its integrated history (which determines its Hubble type and 
bulge-to-disc ratio).

\begin{figure}
\epsfig{file=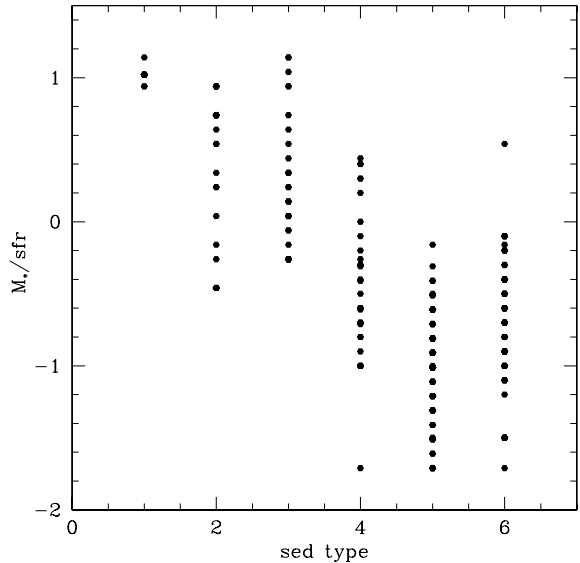,angle=0,width=8cm}
\caption{Plot of $M_*/\dot{M_*}t_0$ versus sed type.}
\end{figure}

\section{Star formation history}

I now pull together estimates of the star-formation history derived from the photometric redshift
methods described above, corrected for the effects of dust extinction, together with other infrared
and submillimetre estimates which should be independent of dust extinction.

\subsection{Estimates from HDF}
In the case of the HDF data the extinction at 2800 $\AA$ is derived using Milky Way grain properties
and using the evolution of $<A_V(z)>$ derived in section 2.2.  The net correction, $10^{0.4 A_{2800}}$,
ranges from 2.1 at z= 0, 2.9 at z =1, to 1.7 at z= 2, 1.3 at z = 3 and 1.16 at z =4.  
Most estimates of dust correction factors to
be applied to high redshift star-forming galaxies range from 2-7 (Meurer et al 1997, 1999,
Pettini et al 1998, Steidel et al 1999).  Tresse and Maddox (1998) derive a correction factor of 
2.5 from a comparison of H$\alpha$ and optical data for z = 0.3 galaxies from the CFRS survey.
The corrections I have applied at
at z = 3 and 4 are much smaller than those assumed by Steidel et al (1999).  The latter are based on
an assumed E(B-V) = 0.15, derived from the (G-R) colour distribution, and the assumption of
a Calzetti (1997) extinction law.  As discussed in the previous section, it is clearly very risky to 
estimate dust extinction from a single colour.
Some doubts on the validity of the Calzetti extinction law have been raised by Rowan-Robinson (2003).

\subsection{New 15 and 850 $\mu$m estimate}
Rowan-Robinson et al (1997) gave an estimate for the star formation rate based on ISO detections
(mainly 15 $\mu$m) of HDF-N galaxies.  Aussel et al (1998) have given a reanalysis of these data
using a wavelets method, which confirms the reality of virtually all the detections used by 
Rowan-Robinson (1997).  There is also reasonably good agreement of the 15
$\mu$m fluxes used in the two studies.  Only source 3 in Table 1 of Rowan-Robinson et al (1997) 
is not confirmed
as a starburst galaxy.  Source 2 is not detected by Aussel et al (1998) but is confirmed as
a starburst galaxy by Richards et al (1998).  


\begin{figure}
\epsfig{file=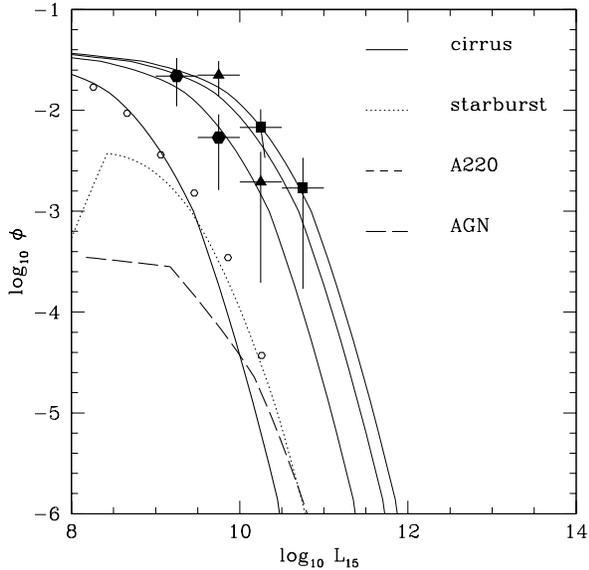,angle=0,width=8cm}
\caption{Luminosity function at 15 $\mu$m for different components from the 
model of Rowan-Robinson (2001), 
compared with revised estimate from the ISO survey of the HDF (filled circles: z = 0.4-0.7, 
triangles: z =
0.7-1.0, squares: z = 1.0-1.3), and the local estimate of Xu et al 
(2000, open circles).}
\end{figure}

Using the noise contours provided by Aussel et al (1998), I have used a 13.5 sq arcmin area
of HDF-N and the Hubble Flanking Fields defined by 7 $\tau_w >$ 150 mJy to estimate the 
15 $\mu$m luminosity function in different redshift bins using the Aussel et al
15 $\mu$m fluxes.   There are 22 ISO 15 $\mu$m 
sources within this area, above this flux-limit.  Spectroscopic redshifts are available
in the Hawaii catalogue for 17 of these, and I have used my photometric redshift
estimates (see section 2.1 above) for the remainder.  
Figure 26 shows the 15 $\mu$m luminosity functions for redshift bins 0.4-0.7, 0.7-1.0,
and 1.0-1.3.  The zero redshift luminosity function of Rowan-Robinson (2001) is also shown
together with the locally observed estimates by Xu et al (2000).  
The luminosity function is also shown shifted by an appropriate pure luminosity evolution for each of 
the 3 redshift bins.  From these shifts I 
estimate  the star-formation rates in these redshift bins.  These estimates based on 
luminosity functions are probably more reliable than estimates made by simply adding
all luminosities, as in Rowan-Robinson et al (1997).  I have also applied the same
method to the 850 $\mu$m data HDF-N of Hughes et al (1998), shown in Fig 27.  For comparison,
Barger et al (2000) derived a star formation rate a factor of 3 higher.

\subsection{Global star-formation history}

Figure 28 shows the star formation
history derived either from ultraviolet or H$\alpha$ data with correction for reddening,
or from far infrared or submillimetre data under the assumption that most ultraviolet and 
visible light
from star forming regions is absorbed by dust.  
The curve corresponds to the model used by Rowan-Robinson (2001) to fit multi-wavelength source-counts
and the spectrum of the background radiation.  At z = 0 it is drawn through the local estimate derived
from the uv luminosity-density by Rowan-Robinson (2003).

Haarsma et al (2000) have given estimates of star-formation rates as a function of
redshift based on deep VLA surveys of HDF-N and the HFF.  These show the same steep rise
with redshift as the ISO estimates, but are typically a factor of two higher. 

The different estimates used in Fig 28 agree surprisingly well with each other and with the model.
The estimates derived from HDF-N and -S lie somewhat below the ISO estimates.  
The surveys planned with SIRTF will be able to resolve these discrepancies.
    
\begin{figure}
\epsfig{file=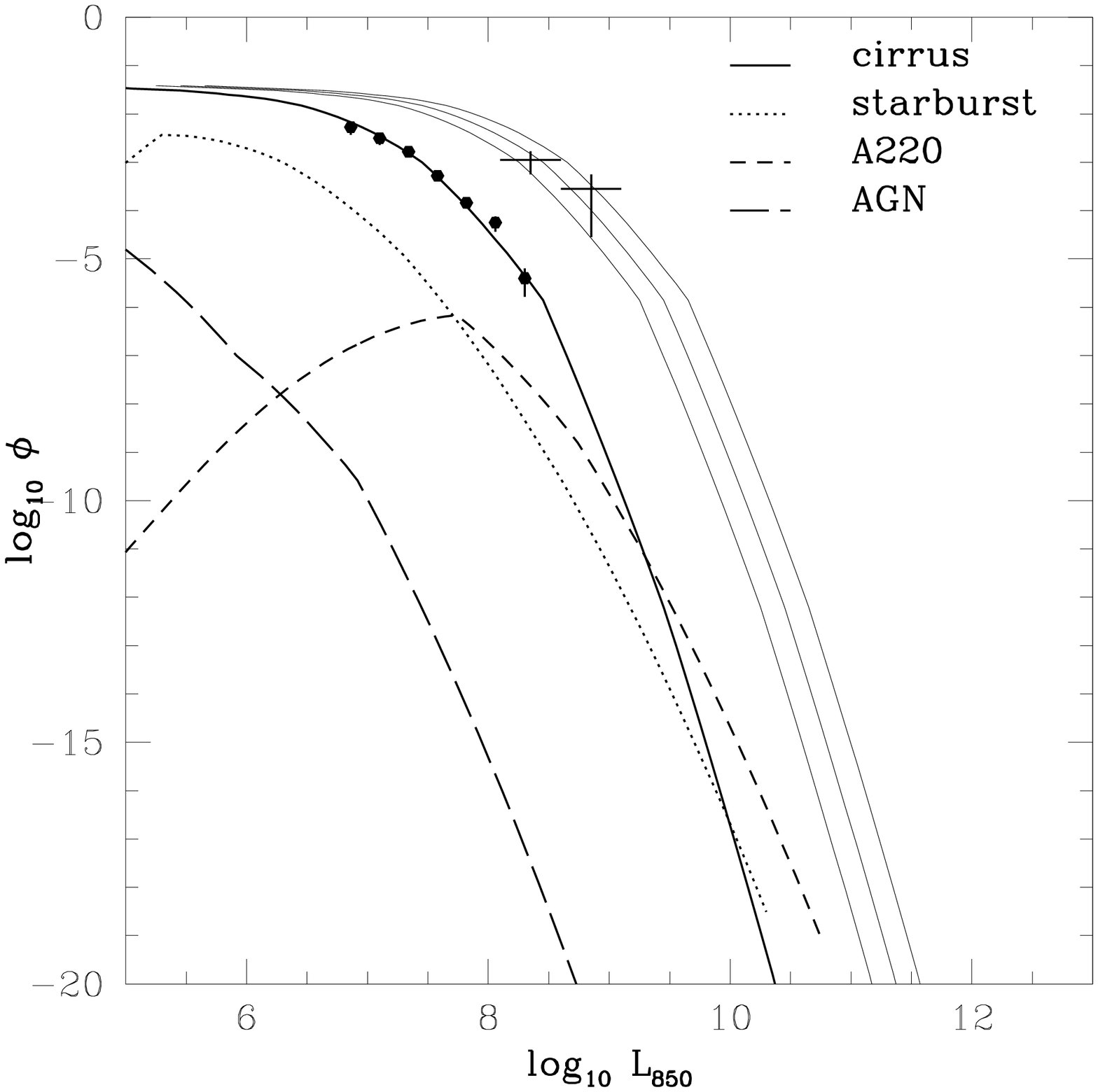,angle=0,width=8cm}
\caption{Luminosity function at 850 $\mu$m, compared with local data of
Dunne et al (2000, filled circles) and with revised estimate from the SCUBA survey of the HDF.}
\end{figure}

\begin{figure}
\epsfig{file=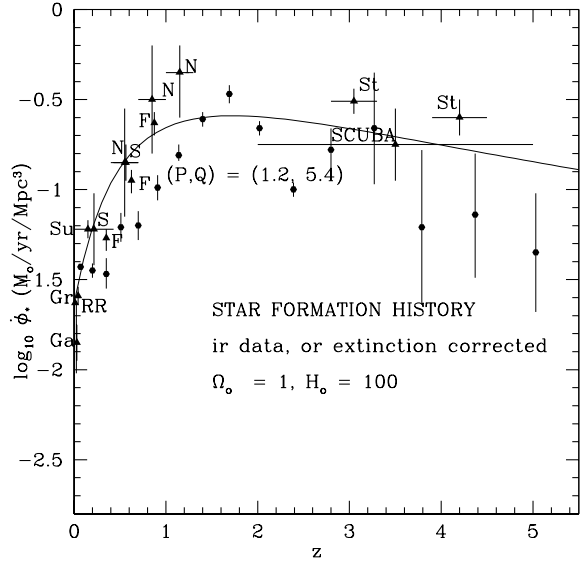,angle=0,width=8cm}
\caption{Star formation history derived for an $\Omega_0$ = 1 universe from infrared (N: 
ISO-HDF-N data (Rowan-Robinson et al 1997) as reanalysed in the present paper, 
S: ISO-HDF-S data of Mann et al (2001), 
F: Flores et al 1999 ISO survey of CFRS field), and submillimeter
(SCUBA, 850 $\mu$m: Hughes et al 1998 data, as reanalysed in the present paper) data, 
from ultraviolet data with
correction for effects of dust, from Gallego et al (1995, Ga), Sullivan et al (2001, Su),
 both corrected by a factor 2 for dust extinction, Gronwall (1998, Gr), Rowan-Robinson (2003, RR) and Steidel
et al (1998, St) (triangles) and from the 
analysis of photometric redshifts of HDF galaxies given in section
3, corrected for dust extinction at 2800 $\AA$ 
as described in section 2.2.  
The model shown is chosen to fit the far infrared and submm counts
(Rowan-Robinson 2001),
 (P,Q) = (1.2, 5.4), and has been drawn through a local, z = 0, value for 
$log_{10} \dot{\phi_{*}}$ = -1.63 (taken from the analysis of local galaxies with uv and far ir data by 
Rowan-Robinson (2003)).} 
\end{figure}

\begin{figure}
\epsfig{file=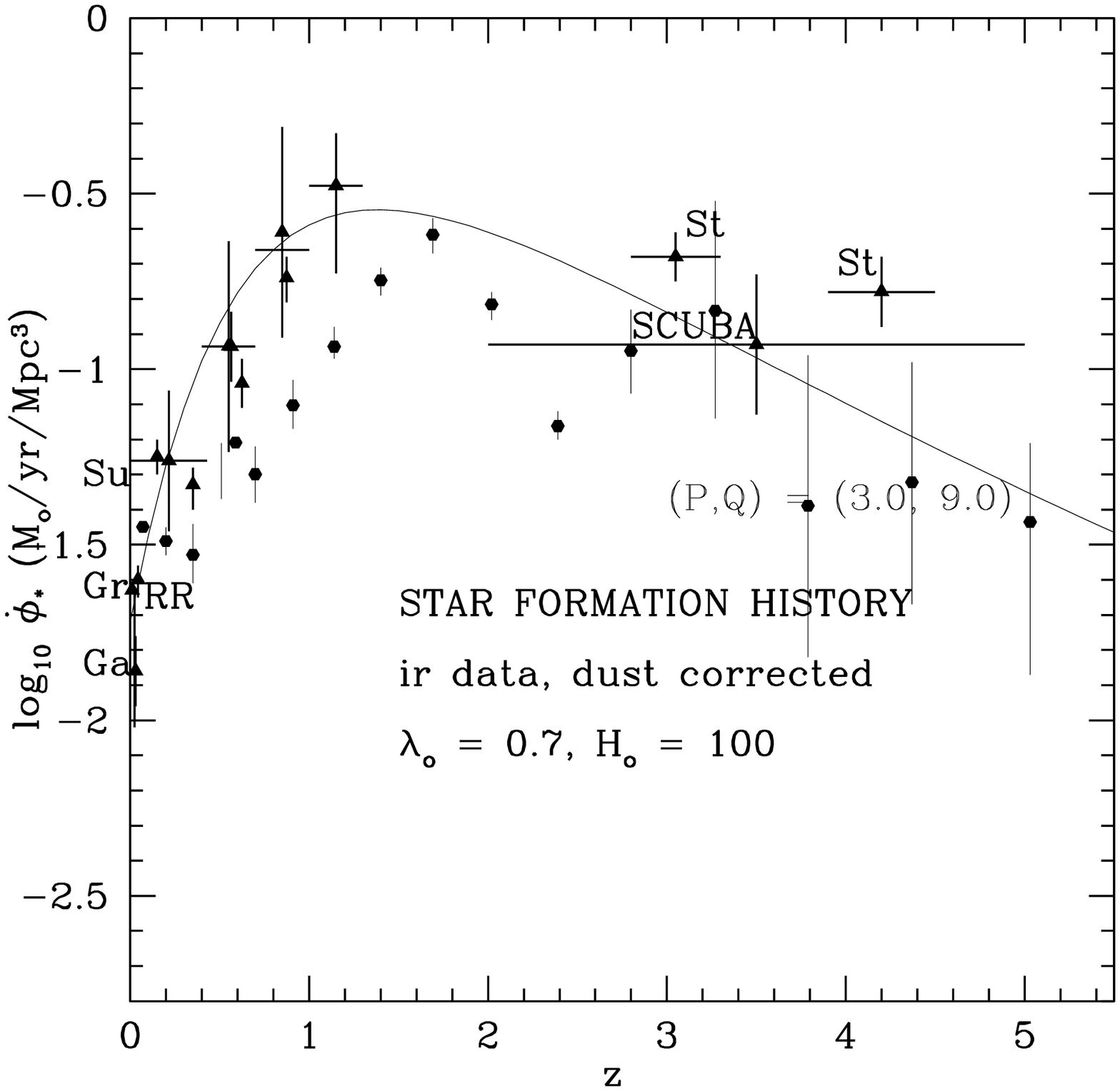,angle=0,width=8cm}
\caption{Star formation history derived from infrared data, 
from ultraviolet data with
correction for effects of dust, and from the 
analysis of photometric redshifts of HDF galaxies given in section
3 for an $\Omega_o = 0.3, \lambda_o = 0.7, H_o$ cosmology.  
The model shown is chosen to fit the far infrared and submm counts
(Rowan-Robinson 2001),
 (P,Q) = (3.0, 9.0), and have been drawn through a local, z = 0, value for 
$log_{10} \dot{\phi_{*}}$ = -1.63 (Rowan-Robinson 2003). }
\end{figure}

It is relatively straightforward to transform the star-formation as a function of redshift
to a different cosmological model.  Figure 29 shows the same data as Fig 28 transformed to a model
with curvature constant k = 0, $\lambda_o$ = 0.7, together with the best fit star-formation
history used in this model by
Rowan-Robinson (2001) to fit source-counts and background radiation from optical to submm
wavelengths.  Both model history and observations have to decline at higher redshifts in this
cosmological model because of the greater volume element available per redshift interval.

\section{Conclusions}

(1) I have investigated the problem of aliasing for photometric redshift estimates and find that
it is a serious problem if less than 4 photometric bands are used.  Where possible I have compared
my photometric redshift estimates with those from other studies. 

(2) With reasonable restrictions, it is possible to determine the dust extinction as well as the 
photometric redshift, provided 5 or more photometric bands are available.  
The expected evolution of  $<A_V(z)>$ with redshift is seen.
The extinction is higher than locally at z= 0.5-1.5, and lower at z $>$ 2.

(3) Deconvolving uv-to-ir seds into an old star and young star component allows determination
of $M_*$ and $\dot{M}_*$ for each galaxy, as well as $z_{phot}$ and $A_V$, provided there are a 
reasonable number of photometric bands available.  The expected trend of b = $\dot{M}_* /<\dot{M}_*>$
increasing to the past is seen.  However there is a great deal of scatter in the relation
between b and uv sed type, showing that the recent star-formation history is not very well
correlated with the long-term history of a galaxy.

(4) I have calculated the 2800 $\AA$ luminosity function and star-formation rate for a large sample 
of HDF-N (2490) and HDF-S (28719) galaxies, using photometric redshifts, 
for the redshift range 0.2-5.  The results agree well with those from a variety of other uv
estimates.  The luminosity functions are consistent with a steep faint-end slope ($\alpha$ = 1.6)
at all redshifts.

\section{Acknowledgements}
I thank Seb Oliver, Thomas Babbedge, and Maria Polletta for helpful discussions, Daniel Le Borgne for 
supplying his photometric redshift catalogue, and an anonyous referee for helpful comments.

\begin{table*}
\caption{Proportions of different galaxy sed types in HDF-N as a function of redshift}
\begin{tabular}{llllllllllllll}
& & & & & & & & & & & & & \\
z = & 0.1 & 0.3 & 0.5 & 0.7 & 0.9 & 1.1 & 1.3 & 1.5 & 1.7 & 1.9 & 2.1 & 2.3 & 2.5 \\
\hline
& & & & & & & & & & & & & \\
$N_{gal}, (B\leq 29.0)$ = & 17 & 62 & 90 & 88 & 107 & 111 & 172 & 172 & 174 & 897 & 124 & 161 & 56 \\
\hline
& & & & & & & & & & & & & \\
$\%$ E & 11.8 & 14.5 & 4.4 & 9.1 & 11.2 & 3.6 & 0.6 & 0.0 & 0.6 & 0.0 & 0.0 & 0.6 & 5.4 \\
& & & & & & & & & & & & & \\
$\%$ Sab & 0 & 3.2 & 3.3 & 0 & 2.8 & 5.4 & 2.9 & 6.4 & 1.7 & 11.5 & 18.5 & 4.3 & 1.8 \\
& & & & & & & & & & & & & \\
$\%$ Sbc & 11.8 & 3.2 & 4.4 & 1.1 & 6.5 & 1.8 & 0.6 & 1.2 & 3.4 & 3.4 & 4.0 & 1.2 & 1.8 \\
& & & & & & & & & & & & & \\
$\%$ Scd & 41.2 & 21.0 & 17.8 & 18.2 & 7.5 & 21.6 & 5.2 & 6.4 & 8.0 & 6.9 & 4.8 & 6.8 & 8.9 \\
& & & & & & & & & & & & & \\
$\%$ Sdm & 11.8 & 17.8& 31.1 & 46.6 & 19.6 & 37.8 & 19.2 & 6.4 & 13.8 & 42.5 & 46.0 & 6.2 & 5.4\\
& & & & & & & & & & & & & \\
$\%$ starburst & 23.5 & 37.1 & 38.9 & 25.0 & 52.3 & 29.7 & 71.5 & 79.7 & 72.4 & 35.6 & 26.6 & 80.7 & 76.8 \\
\end{tabular}
\end{table*}


\begin{table*}
\caption{Objects in HDF-N for which at least one photometric redshift method has a problem.
Columns give name, Fernandez-Soto (FLY) et al (1999) number, spectroscopic redshift,
and photometric redshifts given by FLY(1999), FLY(2001), le Borgne and Rocca-Volmerange (2002),
present work template method (RR,T), template method + arbitrary $A_V$ (RR,TAV), two
component method (RR,C), two component method + arbitrary $A_V$.  Bottom line gives total
number of problem sources for each method,
defined as $\Delta = |log_{10} ((1+z_{ph})/(1+z_{sp}))| > 0.10$.
}
\begin{tabular}{llllllllll}
& & & & & & & & & \\
name & FLY & $lg_{10}$ & $ lg_{10}$ & $lg_{10}$ & $lg_{10}$ & $lg_{10}$ &
$lg_{10}$ & $lg_{10}$ & $lg_{10}$\\
 & no. & $(1+z_{sp})$ & $(1+z_{FLY})$ & $(1+z_{FLY01})$ & $(1+z_{BR})$ &
$(1+z_{RR,T})$ & $(1+$ & $(1+z_{RR,C})$ &
$(1+$\\
& & & & & & & $z_{RR,TAV})$ & & $z_{RR,CAV})$ \\
& & & & & & & & & \\
3-875.0 & 48 & 0.484 & 0.483 & 0.519 & 0.00 & 0.47 & 0.47 & 0.47 & 0.46 \\
4-332.1 & 315 & 0.225 & 0.326 & 0.238 & 0.00 & 0.28 & 0.25 & 0.26 & 0.25 \\
3-550.1 & 334 & 0.577 & 0.435 & 0.000 & 0.00 & 0.48 & 0.53 & 0.48 & 0.54 \\
4-289.0 & 444 & 0.599 & 0.483 & 0.567 & 0.000 & 0.55 & 0.53 & 0.55 & 0.56 \\
4-928.1 & 466 & 0.304 & 0.283 & 0.286 & 0.290 & 0.29 & 0.20 & 0.27 & 0.15 \\
4-445.0 & 517 & 0.544 & 0.365 & 0.540 & 0.447 & 0.48 & 0.44 & 0.42 & 0.43 \\
3-243.0 & 568 & 0.627 & 0.615 & 0.653 & 0.630 & 0.63 & 0.63 & 0.63 & 0.47 \\
4-254.0 & 653 & 0.279 & 0.204 & 0.201 & 0.230 & 0.18 & 0.13 & 0.21 & 0.23 \\
4-52.111 & 687 & 0.595 & 0.107 & 0.100 & 0.097 & 0.63 & 0.63 & 0.63 & 0.46 \\
4-639.1 & 702 & 0.555 & 0.000 & 0.531 & 0.000 & 0.57 & 0.57 & 0.58 & 0.58 \\
4-946.0 & 716 & 0.289 & 0.225 & 0.230 & 0.176 & 0.26 & 0.19 & 0.21 & 0.21 \\
2-585.1111 & 1016 & 0.474 & 0.422 & 0.422 & 0.362 & 0.49 & 0.44 & 0.44 & 0.43 \\
2-251.0 & 1018 & 0.293 & 0.265 & 0.267 & 0.190 &0.27 & 0.20 & 0.21 & 0.19 \\
2-449.1 & 1044 & 0.478 & 0.017 & 0.543 & 0.000 & 0.49 & 0.49 & 0.49 & 0.49 \\
2-82.1 & 1062 & 0.514 & 0.435 & 0.004 & 0.512 & 0.53 & 0.54 & 0.54 & 0.56 \\
& & & & & & & & & \\
number & with & $\Delta > 0.10$ & 7 & 3 & 10 & 0 & 3 & 1 & 5 \\
& & & & \\
\end{tabular}
\end{table*}


\begin{table*}
\caption{Galaxies in HDF-N with $z_{phot} >$ 5.}
\begin{tabular}{lllllll}
name & I & $z_{phot}$ & $z_{sp}$ & $z_{FLY}$ & $n_{type}$ & $n_{band}$ \\
& & & & & &  \\
 3-951.1   & 25.83 & 6.08 & 5.339 & 5.72 &  4 &  3 \\
 4-625.1   & 25.19 & 5.17 & 4.585 & 4.52 &  4 &  4 \\
 4-314.0   & 27.38 & 6.08 & 0.000 & 4.76 &  3 &  2 \\
 4-473.0   & 27.95 & 6.08 & 5.607 & 5.64 &  1 &  2 \\
 4-842.0   & 28.02 & 6.08 & 0.000 & 4.64 &  1 &  2 \\
 4-148.0   & 28.78 & 5.46 & 0.000 & 1.24 &  4 &  2 \\
 4-169.0   & 26.61 & 6.08 & 0.000 & 5.64 &  3 &  3 \\
 3-516.0   & 27.22 & 5.92 & 0.000 & 4.76 &  1 &  2 \\
 3-489.0   & 27.06 & 6.08 & 0.000 & 4.76 &  3 &  2 \\
 3-457.0   & 27.33 & 6.08 & 0.000 & 5.48 &  3 &  2 \\
 3-387.0   & 26.28 & 5.31 & 0.000 & 0.84 &  2 &  4 \\
 3-324.0   & 28.61 & 5.46 & 0.000 & 5.20 &  3 &  2 \\
 3-255.0   & 27.50 & 5.92 & 0.000 & 4.92 &  4 &  2 \\
 3-242.0   & 27.72 & 5.31 & 0.000 & 0.68 &  3 &  2 \\
 3-228.0   & 27.25 & 5.76 & 0.000 & 5.04 &  1 &  2 \\
 3-153.0   & 27.20 & 6.08 & 0.000 & 5.32 &  4 &  2 \\
 4-799.0   & 27.67 & 6.08 & 0.000 & 4.84 &  1 &  2 \\
 4-884.0   & 28.37 & 5.46 & 0.000 & 4.40 &  5 &  2 \\
 4-310.0   & 28.42 & 6.08 & 0.000 & 4.84 &  1 &  2 \\
 3-44.0    & 27.72 & 5.76 & 0.000 & 4.80 &  1 &  2 \\
 2-282.0   & 26.14 & 5.76 & 0.000 & 4.36 &  2 &  3 \\
 2-892.0   & 26.50 & 5.76 & 0.000 & 4.72 &  4 &  2 \\
 2-156.0   & 27.62 & 5.46 & 0.000 & 0.72 &  4 &  2 \\
 2-925.0   & 28.26 & 5.31 & 0.000 & 4.60 &  3 &  2 \\
 2-364.0   & 28.22 & 6.08 & 0.000 & 4.80 &  1 &  2 \\
2-490.0    & 28.60 & 5.03 & 0.000 & 0.00 &  4 &  2 \\
2-882.0    & 27.35 & 5.31 & 0.000 & 0.00 &  4 &  2 \\
3-839.0    & 24.66 & 5.92 & 0.000 & 0.00 &  5 &  3 \\
3-874.0    & 27.97 & 5.76 & 0.000 & 0.00 &  3 &  2 \\
3-951.0    & 25.58 & 5.76 & 0.000 & 0.00 &  5 &  2 \\
3-965.1112 & 25.53 & 5.31 & 0.000 & 0.00 &  4 &  3 \\
4-625.2    & 27.49 & 5.61 & 0.000 & 0.00 &  1 &  2 \\
4-683.0    & 29.72 & 5.03 & 0.000 & 0.00 &  1 &  2 \\
\end{tabular}
\end{table*}


\section{Appendix: Star formation history model}

For a simple closed box galaxy model with constant yield and instantaneous recycling, 
in which all the heavy elements in the interstellar medium
are assumed to be in the form of dust, the mass of gas $M_{gas}(t)$, stars $M_{*}(t)$, heavy elements $M_{Z}$, and
dust $M_{dust}(t)$, satisfy

\medskip

$dM_*/dt = \dot{\phi}_* = - dM_{gas}/dt$,	(A.1)

\medskip

$M_Z = M_{dust} = k M_*$,			(A.2)

\medskip

$\dot{M}_{dust} =( k - Z) \dot{\phi}_*$,	(A.3)

\medskip

$M_{dust} = Z M_{gas}$,				(A.4)

\medskip

where k is the yield, Z is the metallicity of the ism, and $\dot{\phi}_*$ is the star-formation rate.

Substituting (A.4) in (A.3) and using (A.1), it follows that

\medskip

$M_{gas} \dot{Z} = - k \dot{M}_{gas}$

\medskip

and hence (Pei et al 1999) that

\medskip

$Z = k ln(M_{gas}(0)/M_{gas}(t))$		(A.5)

\medskip

and

$M_{dust}(t) = k M_{gas}(t) ln(M_{gas}(0)/M_{gas}(t))$.	(A.6)

\medskip

Rowan-Robinson (2001) has shown that star-formation histories of the form

\medskip

$\dot{\phi}_* = \alpha t^m e^{-\nu t}$     (A.7)

\medskip

provide an extremely effective fit to source-counts and redshift distributions from optical to submillimetre
wavelengths, with the assumption that characteristic luminosities in galaxies trace
$\dot{\phi}_*(t)$.  In such a model

\medskip

$M_*(t) = \alpha_0 M_{gas}(0) \gamma(m+1,\nu t)$	(A.8)

$= M_{gas}(0) - M_{gas}(t)$

\medskip

where the incomplete gamma function

\medskip

 $\gamma(m+1, y) = \int_0^y x^m e^{-x} dx$, and

\medskip

$\alpha = \alpha_0 \nu^{m+1} M_{gas}(0)$.	(A.9)

\medskip

There is then a critical value of $\alpha_0$, $\alpha_{0,c} = 1/\Gamma(m+1)$, such that the gas runs
out in the galaxy if $\alpha_0 > \alpha_{0,c}$.  This, together with the fact that the star-formation
rate starts off as zero at t = 0, rather than at its maximum value, makes models of type (A.7)
more versatile and realistic than the usually assumed exponentially declining star-formation rate
(Bruzual and Charlot 1993, Pei et al 1999, Fioc and Rocca-Volmerange 2000).  Equations (A.6) and (A.8) have
been used to provide the model curves in Figs 20 and 24 of this paper, with the assumption that
the visual extinction

\medskip

$A_V(t) = ( 3 Q_V/4 a \rho_{gr}) M_{dust}(t) / \pi R_{gal}^2$,	(A.10)

\medskip

where $R_{gal}$, a, $\rho_{gr}$, $Q_V$ are characteristic values for the galaxy radius, grain radius,  
density of grain material, and V-band extinction efficiency.  In practice, having specified m, $\nu$
($ = P, Q/t_0$ in the notation of Rowan-Robinson 2001), $\alpha_0$ and the parameters in eqn (A.10)
are chosen to give a gas fraction of 10$\%$ and an $A_V$ of 0.3 (Rowan-Robinson 2003) at the present 
epoch.

To fully match the properties of present day galaxies it is necessary to modify eqn (A.1) to
take account of both infall of pristine intergalactic gas into galaxies, and outflow of enriched
material in supernova-driven winds.  Some aspects of inflow can be modelled by extending (A.7)
to include density evolution, to match the effects of galaxy merging (Rowan-Robinson 2001).
King and Rowan-Robinson (2003) have used such models to improve the fit to optical and near infrared
galaxy counts, and shown that the best-fitting models of this type give evolving luminosity
functions very similar to those found in heirarchical simulations.  

If we take the space-density of galaxies, as in King and Rowan-Robinson (2003) to be modified by a factor
$\rho(z) = (1+z)^n$, we then need to think of each galaxy at z = 0 broken up into $\rho(z)$ pieces
at z, so $R_{gal}(z) = R_{gal}(0) \rho(z)^{-1/3}$ and then

\medskip

$A_V(t) = ( 3 Q_V/4 a \rho_{gr}) M_{dust}(t) \rho(z)^{-1/3}/ \pi R_{gal}^2$.	(A.11)

\medskip

The King and Rowan-Robinson (2003) model for $\Omega_0$ = 1 has been included in Fig 20, with (A.7) modified
to replace m by $P-2n/3$.

\end{document}